\documentclass[a4paper,fleqn,usenatbib]{mnras}
\usepackage[T1]{fontenc}
\usepackage{ae,aecompl}
\usepackage{graphicx}   
\usepackage{amsmath}    
\usepackage{amssymb}    
\usepackage{times}
\usepackage{xspace}

\newcommand{\apm}{APM~08279+5255\xspace}
\newcommand{\pds}{PDS~456\xspace}
\newcommand{\hh}{1H~0707-495\xspace}
\newcommand{\xmm}{{\it XMM-Newton}\xspace}
\newcommand{\suzaku}{{\it Suzaku}\xspace}
\newcommand{\nustar}{{\it NuSTAR}\xspace}
\newcommand{\chandra}{{\it Chandra}\xspace}
\newcommand{\feka}{Fe\,K$\alpha$\xspace}

\newcommand{\fexxv}{Fe\,\textsc{xxv}\xspace}
\newcommand{\fexxvi}{Fe\,\textsc{xxvi}\xspace}
\newcommand{\civ}{C\,\textsc{iv}\xspace}

\newcommand{\xstar}{{\sc xstar}\xspace}
\newcommand{\monaco}{{\sc monaco}\xspace}

\newcommand{\msun}{M$_\odot$\xspace}

\title[A disc wind of \apm]
{Revisiting the extremely fast disc wind in a gravitationally lensed quasar \apm}

\author[K. Hagino et al.]
{Kouichi Hagino,$^1$\thanks{E-mail: hagino@astro.isas.jaxa.jp}
Chris Done,$^{1,2}$
Hirokazu Odaka,$^3$
Shin Watanabe,$^{1,4}$
\newauthor
and Tadayuki Takahashi$^{1,4}$\\
$^1$ Institute of Space and Astronautical Science (ISAS), Japan Aerospace Exploration Agency (JAXA), 3-1-1 Yoshinodai, Chuo, Sagamihara,\\ Kanagawa 252-5210, Japan\\
$^2$ Department of Physics, University of Durham, South Road, Durham DH1 3LE, UK\\
$^3$ KIPAC, Stanford University, 452 Lomita Mall, Stanford, CA 94305, USA\\
$^4$ Department of Physics, University of Tokyo, 7-3-1 Hongo, Bunkyo, Tokyo 113-0033, Japan\\
}

\date{Accepted XXX. Received YYY; in original form ZZZ}

\pubyear{2017}

\begin{document}
\label{firstpage}
\pagerange{\pageref{firstpage}--\pageref{lastpage}}
\maketitle

\begin{abstract}
The gravitationally lensed quasar \apm has the fastest claimed wind
from any AGN, with velocities of 0.6--0.7$c$, requiring magnetic
acceleration as special relativisitic effects limit all radiatively
driven winds to v<0.3--0.5$c$.  However, this extreme velocity derives
from interpreting both the narrow and broad absorption features in the
X-ray spectrum as iron absorption lines.  The classic ultrafast
outflow source \pds also shows similar absorption systems, but here
the higher energy, broader feature is generally interpreted as an
absorption edge. We reanalyse all the spectra from \apm using a full
3-dimensional Monte Carlo radiative transfer disc wind model for the
ionised wind at 0.1--0.2$c$, together with complex absorption from
lower ionisation material, and find that this is a better description
of the data. Thus there is no strong requirement for outflow
velocities beyond $0.2c$, which can be powered by radiation
driving. We show that UV line driving is especially likely given the
spectral energy distribution of this source which is intrinsically UV
bright and X-ray weak.  While the peak of this emission is
unobservable, it must be luminous enough to power the observed hot
dust, favouring at least moderate black hole spin. 

\end{abstract}

\begin{keywords}
black hole physics -- radiative transfer -- galaxies: active -- galaxies: individual: \apm -- X-rays: galaxies.
\end{keywords}

\section{Introduction}
Accretion disc winds from Active Galactic Nuclei (AGN) are thought
to play an important role in the evolution of the supermassive black
holes and their host galaxies. Winds with a kinetic power which is
only 5 per cent of the Eddington luminosity, $L_{\rm Edd}$, can quench
star formation in the bulge by sweeping away the gas reservoir, and
quantitatively reproduce the observed relation between the properties
of the black hole and the galaxies \citep[e.g.,][]{King2010}.

The most powerful winds in AGN have recently been revealed by 
X-ray observations showing ultra-fast outflows (UFOs)
\citep[e.g.,][]{Chartas2002,Reeves2003,Pounds2003,Pounds2003a,
Tombesi2010,Gofford2013}.
These are seen as absorption lines of \fexxv and/or
\fexxvi ions in the X-ray band, blueshifted by more than 10,000~km/s
i.e. $0.03c$ \citep{Tombesi2010}. Such features were seen in up to
35\% of local AGN (\citealt{Tombesi2010} but see \citealt{Laha2014} and
\citealt{Tombesi2014}). The most convincing have large column density
of $\sim 10^{23}$~cm$^{-2}$ and fast velocity of $\sim0.1c$, where the
associated kinetic power is estimated to be high enough to play a key
role in the co-evolution of the black holes and galaxies.

In spite of its importance, the physical properties of UFOs are
not fully understood. One of the major uncertainties is how the
outflows are launched and accelerated. Continuum driven winds require
$L\gtrsim L_{\rm Edd}$, while many of the sources with detected
(though sometimes controversial) UFOs have $L\sim 0.1L_{\rm Edd}$
\cite[e.g., IC 4329A, Mkn 509, Akn 120: ][]{Laha2014,Tombesi2014}.
UV line driving only works if the material 
has substantial UV opacity i.e. is 
not highly ionised. Strong X-ray illumination
will overionise the material, and shielding the gas \citep[e.g.,][]{Murray1998,Proga2004} is not easy as X-rays can scatter
around the shield \citep{Higginbottom2014}. Thermal winds, driven by
the pressure gradient of X-ray heated gas, have much smaller
velocities as they are launched at fairly large distances from the
source, where material heated to the Compton temperature is unbound
\citep{Begelman1983}. This only leaves magnetic driving, which
depends on the (unknown) field geometry, so no predictions are
possible \citep{Proga2000}.

Without a physical mechanism, most current studies of UFOs have
concentrated on constraining the physical properties of outflow.
However, it is difficult to self-consistently model the emission and
absorption from the UFOs since the wind geometry is probably not
spherical \citep{Elvis2000,Proga2004,Risaliti2010}. Such asymmetric
geometries require Monte Carlo radiative
transfer calculations to derive the emission and absorption self
consistently. 
Such simulations were performed by \cite{Sim2008,Sim2010a,Sim2010b},
but were used in detailed modelling of only two individual sources, PDS 456 \citep{Reeves2014} and
PG 1211+143 \citep{Sim2010a}. 

In our previous work \citep{Hagino2015}, we developed a new 3D Monte
Carlo simulation code for accretion disc winds in order to match to
observational data from UFOs.  Our code can calculate radiative
transfer in H- and He-like ions in a realistic accretion disc wind geometry.
We applied this simulation to all the \suzaku spectra of \pds, and successfully
reproduced the changing UFO properties seen in this source by moderate
changes in the velocity ($0.2\textrm{--}0.3c$) and the angle to the line of sight
of a disc wind. However, the main new aspect of this paper was that it
re-assessed the possibility that the outflow was a UV line driven disc
wind as this object has $L\sim L_{\rm Edd}$, with a spectral energy
distribution which peaks in the UV and is X-ray weak. A more
favourable set of circumstances for UV line driving (helped by
radiation pressure as $L\sim L_{\rm Edd}$) is hard to imagine. The outflow velocity is also
characteristic of UV line driven disc winds \citep{Proga2004,
Risaliti2010,Nomura2013,Nomura2016}, as is the fluctuation
behaviour about a steady state structure \citep{Proga2004}.  The
observed ionisation state is far too high for UV line driving, but the
acceleration could take place in much lower ionisation material close
to the disc, which becomes ionised and enters the line of sight only
when it has lifted high enough to be ionised by illumination from the hotter inner disc and
X-ray source \citep{Hagino2015}.

Our wind model was also applied to a putative broad iron line feature in \hh
\citep{Hagino2016}. The characteristic sharp drop at $\sim7$~keV in the X-ray
spectra of this source had been interpreted as the blue end of an extremely smeared
disc reflection spectrum, requiring maximal black hole spin and a very low height of the
point-like corona. Instead, our wind model successfully reproduced all the \xmm observations
of \hh (and the \nustar data) without any constrains on black hole spin, for a wind
velocity of $v=0.2c$ and a mass outflow rate of 
$\dot{M}_{\rm wind}/\dot{M}_{\rm Edd}=0.2$. Interestingly, a closer look at the
fit residuals suggest that the P Cygni iron emission line from the wind underpredicts the observed
iron-K line emission. This could indicate that the wind has a larger opening angle than 
the model assumption of $\Omega/2\pi=0.15$, as expected from a highly super Eddington source \cite{Done2016}.

However, neither UV line driving nor continuum radiation driving can launch a
wind with an outflow of velocity $v\sim0.7c$ found in a
gravitationally lensed quasar \apm
\citep{Chartas2002,Saez2009,Chartas2009,Saez2011} since radiative driving
can accelerate only up to $\sim0.3\textrm{--}0.4c$ due to radiation drag effects
\citep{Takahashi2015}. Thus, this fast wind is evidence for a 
magnetic driving mechanism \citep{Fukumura2010}.
This source is a high redshift ($z=3.91$) quasar, so that blueshifted
H/He-like Fe lines are seen at $\sim2$~keV. At such an energy, the sensitivities of
current instruments are much better than at $7-8$~keV where the 
absorption lines of UFOs at low redshift are observed.
This enables the UFO signatures to be detected even though the X-ray
flux is roughly one order of magnitude lower than \pds.

In this paper, we use our Monte Carlo wind code to fit the multi-epoch X-ray data
from \apm, to critically reassess whether the extremely fast velocities are required. We
find we get good fits with a wind at 0.1--0.2$c$. We reassess the launch mechanism for the wind 
from the broad band spectra energy distribution. While the overall Eddington fraction of the accretion
flow is poorly constrained due to uncertainties in the magnification from lensing, the shape is UV
bright and X-ray weak, favouring UV line driving. 
We assume a standard cosmology with
$H_0=71$\,km\,s$^{-1}$\,Mpc$^{-1}$, $\Omega_{\rm m}=0.27$ and
$\Omega_\Lambda=0.73$, so that the redshift of the target $z = 3.91$
corresponds to the luminosity distance of $d_{\rm L} =35.5$~Gpc and
co-moving distance of $7.2$~Gpc.

\section{Observational data and comparison with the other strong wind sources}

\apm has been observed by \chandra, \xmm and \suzaku as listed in
Table~\ref{tab:xmmobs}.
We use the same naming convention of \cite{Chartas2009,Saez2011}
except for the first \xmm observation, which was not included in their
analysis due to its short exposure time. We refer to this data as Epoch 0. 

\begin{table*}
\caption{\xmm, \chandra and \suzaku observations of \apm}
\centering
\begin{tabular}{lcclc}
\hline\hline
Name & Observatory & Obs ID & Start Date & Net exposure (ks)$^a$\\
\hline
Epoch 0 & \xmm & 0092800101 & 2001-10-30 & 16.7/16.7/12.3\\
Epoch 1 & \chandra & 2979 & 2002-02-24 & 88.8\\
Epoch 2 & \xmm &  0092800201 & 2002-04-28 & 76.4/77.2/63.2\\
OBS1 &  \suzaku & 701057010 & 2006-10-12 & 102.3/102.3\\
OBS2 &  \suzaku & 701057020 & 2006-11-01 & 102.3/102.3\\
OBS3 &  \suzaku & 701057030 & 2007-03-24 & 117.1/117.2\\
Epoch 3 & \xmm &  0502220201 & 2007-10-06 & 68.0/68.6/39.3\\
Epoch 4 & \xmm &  0502220301 & 2007-10-22 & 75.8/75.8/57.9\\
Epoch 5 & \chandra & 7684 & 2008-01-14 & 88.1\\
\hline
\end{tabular}
\begin{flushleft}
\footnotesize
$^a$ Net exposure time of MOS1/MOS2/PN for \xmm, ACIS for \chandra and FI/BI for \suzaku, respectively.
\end{flushleft}
\label{tab:xmmobs}
\end{table*}%

We processed EPIC-pn and -MOS data and removed dead and hot pixels by
using SAS tasks {\sc epproc} and {\sc emproc} (SAS v.13.5.0)
respectively. Time intervals when background rates of ${\rm
PATTERN}=0$ events at energy $>10$~keV are higher than
0.35~counts~s$^{-1}$ for MOS and 0.4~counts~s$^{-1}$ for pn camera
were removed.  Only events with ${\rm PATTERN}\leq12$ for MOS and
${\rm PATTERN}\leq4$ for pn were considered in the spectral analysis.
The total net exposure times are listed in
Tab.~\ref{tab:xmmobs}. Spectra were extracted from circular regions of
64$\arcsec$ diameter, while background spectra were extracted from
circular regions of the same diameter for pn and annular regions from
100$\arcsec$ to 300$\arcsec$ diameter for MOS in the same chip as the
source regions. We generated the corresponding response matrix (RMF)
and auxiliary response (ARF) files by utilizing {\sc rmfgen} and {\sc
arfgen}. 

\chandra data were reprocessed and extracted using the CIAO tools
{\sc chandra\_repro} and {\sc specextract}. 
Spectra were extracted from
circular regions of 4$\arcmin$ diameter, while background spectra were
extracted from annular regions from 6$\arcmin$ to 30$\arcmin$ diameter.
The spectra were then grouped to obtain a minimum 40 counts in each bin.

We reduced the Suzaku XIS data with standard screening conditions: grade 0, 2, 3, 4
and 6 events were used. Data within 436~s of passage through the South Atlantic
Anomaly, and within an Earth elevation angle $<5^\circ$ and Earth daytime
elevation angles $<20^\circ$ were excluded. Spectra were extracted from circular
regions of 2.9~arcmin diameter, while background spectra were extracted from
annular region from 7.0 to 14.0~arcmin diameter.

\begin{figure}
\centering
\includegraphics[width=1.0\hsize]{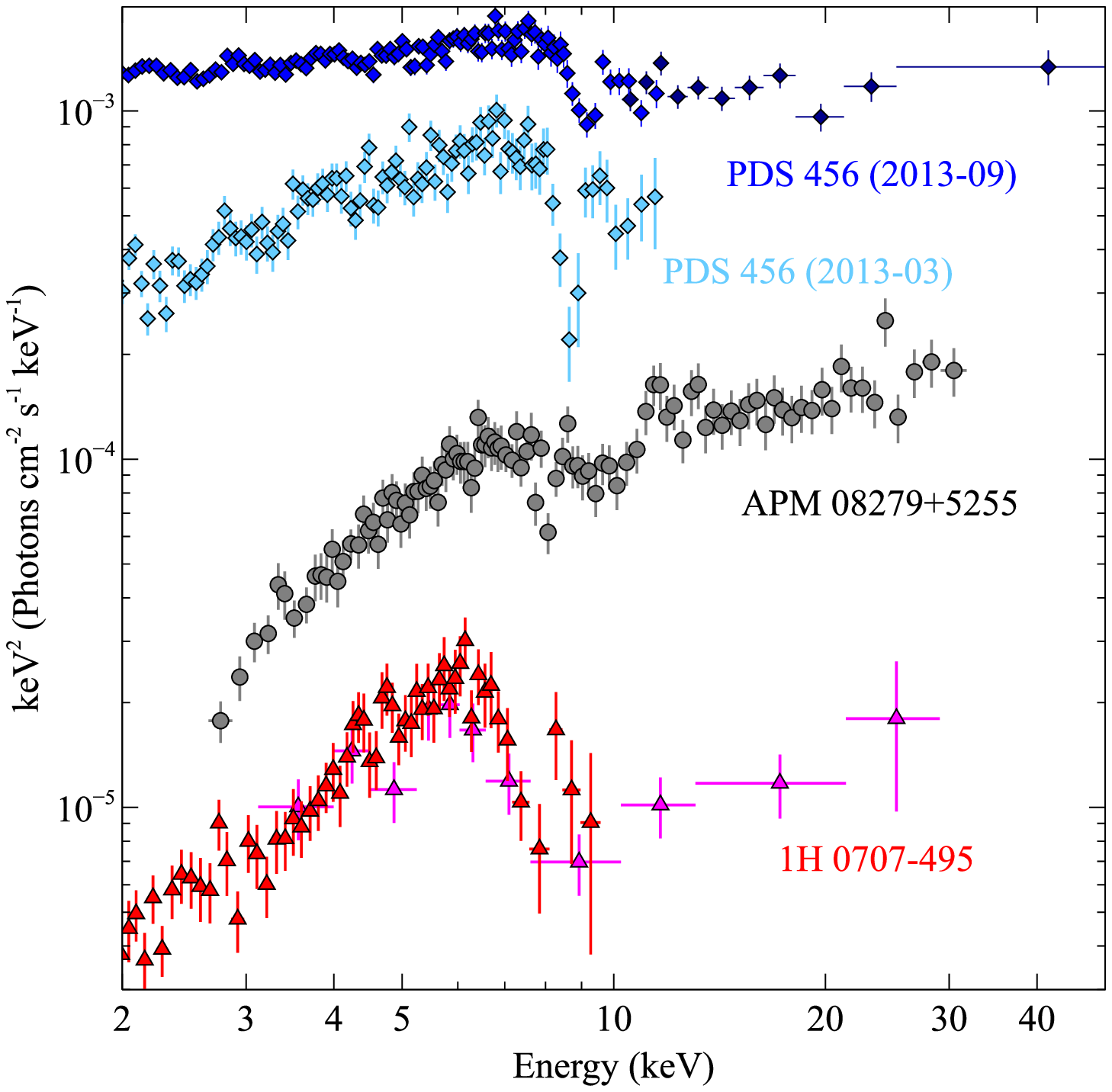}
\caption{ Comparison between the spectra of \pds (blue / dark blue /
  cyan), \hh (red / magenta) and \apm (grey: Epoch1). We show two
  spectra of \pds, one from 2013-09 where there is simultaneous
  \xmm-\nustar data, and one from \suzaku in 2013-03 where the absorption line is
  strongest. The \xmm-\nustar data of \hh are not simultaneous, but they are well matched. 
All spectra are shown in the rest frame, and \hh is scaled down by a factor of 10 for plotting purposes.
All spectra show a narrow absorption line, and the one in \apm is less blueshifted than
that in \pds. All spectra also show evidence for a broader absorption feature at higher energies,
$\sim 10$~keV. It is this feature in \apm which is the evidence for a $0.6c$ wind, yet in \pds and \hh this feature is 
instead interpreted as mainly a photo-electric absorption edge from cool clumps embedded in the wind
which also give rise to the hard 2--6~keV spectrum. 
}
\label{fig:apm_pds}
\end{figure}

Fig.~\ref{fig:apm_pds} shows the strongest absorption line spectrum of
\apm (grey, 2002 \chandra data), compared with the strongest
absorption line states of \pds (cyan, \suzaku 2013-03, ObsID:
707035030) and \hh (red, \xmm 2011, ObsID: 0554710801).  All the
spectra are plotted in the rest frame so the \apm data extend up to
30~keV.  Hence we also extend the \hh and \pds data to higher energies
by including \nustar data.  For \pds, we additionally plot the simultaneous \xmm
(blue, ObsID: 0721010401) and \nustar (dark blue, ObsID: 60002032006)
spectra in 2013-09 \citep{Nardini2015,Matzeu2016}. However, for \hh,
there is no simultaneous observations of \xmm and
\nustar. Nonetheless, we plot the \nustar data (magenta, 60001102004)
as it matches well to the \xmm data in both flux and spectral
shape \citep{Kara2015}.
	
The spectra of these sources shows very similar features in both
continuum and lines. There is a fairly clear, relatively narrow
absorption line, and then a broader absorption feature at higher
energies in all the spectra. In \pds and \hh, these absorption features are
strongest when the 2--5~keV continuum is hardest \citep{Hagino2015,Hagino2016,Matzeu2016},
and the spectrum of \apm is similarly hard. 
It is clear that the narrow absorption lines in \apm
are {\em less} blueshifted than in \pds ($v\sim0.3c$), making it unlikely
that the wind is faster in \apm than in \pds. 

The requirement for extreme velocity in \apm (0.40c for the grey
spectra shown in Fig.~\ref{fig:apm_pds}) comes instead from the broad
component of the absorption. Interpreting this as due to the \feka
line requires material with a large velocity spread, as well as
material with a smaller spread to produce the narrower components of
the line absorption \citep{Chartas2009,Saez2011}. However, as
discussed by these authors, there is an alternative model, where the
broad absorption is from bound-free edges. They showed that the edges
predicted by the same highly ionised material as gives rise to the
line is not completely sufficient to explain the data. However, in
\pds, the broad absorption feature is generally interpreted as a
complex absorption edge, with some contribution from the highly
ionised wind material but with the majority produced in lower
ionisation material which is required to explain the continuum
absorption at lower energies \citep{Hagino2015,Matzeu2016}.  It is
clear that \apm is more absorbed at low energies than even the most
absorbed spectrum of \pds, so it is feasible that it has
stronger edges, producing broad absorption up to $\sim 12$~keV in
these data. The narrow line component is blueshifted to $\sim
7.8$~keV, so requires velocities of $\sim 0.15c$, but here we explore
whether the broader absorption structure at higher energies requires
an additional faster wind or whether they can be produced in the same
absorption structure which gives the hard spectrum below 5~keV.

\section{Monte Carlo simulations of the wind}
\subsection{Code overview}
Our simulation code performs a radiative transfer calculation in a
realistic wind geometry with a Monte Carlo method using \monaco \citep{Odaka2011}.  \monaco
is a general-purpose code for calculating the X-ray spectra from many
astrophysical objects by tracking photon propagation and interaction with
matter.  The interaction position is determined by randomly drawing from 
an exponential distribution with a mean free path of the
interactions, then the photon is absorbed or re-emitted according to
the cross-sections of the interactions.

The physical processes in highly photoionised plasma are already
implemented in the \monaco \citep[see][]{Watanabe2006}.
Photoionisation, photoexcitation, radiative recombination,
de-excitation and Compton scattering by free electrons are taken into
account.  As with the previous work \citep{Hagino2015,Hagino2016},
only H- and He-like ions of Fe and Ni are considered in this work,
which is reasonable assumption for the highly ionised winds like UFOs.

The ionisation structure in the accretion disc wind is calculated by
sequentially running \xstar \citep{Kallman2004}, and then fixed during
the radiative transfer calculation with \monaco.  Ideally, the
radiative transfer simulation and the ionisation structure calculation
should be calculated iteratively.  However, it is not realistic to
repeat the time-consuming Monte Carlo simulations many times, so that
this simplified procedure is adopted.

We use a biconical
geometry, which is often used for studying the radiative transfer in
the accretion disc wind
\citep{Shlosman1993,Knigge1995,Sim2008,Sim2010a}.  This geometry is
described by 3 parameters: the solid angle $\Omega$ (or the covering
factor $\Omega/4\pi$), the minimum radius $R_{\rm min}$ and the inner
angle $\theta_{\rm min}$.  The radial velocity follows an extension of
the CAK velocity law \citep{Castor1975}, parametrized by the initial
velocity $v_0$, the terminal velocity $v_\infty$ and the acceleration
index $\beta$. The rotational velocity and the density are determined
by conservation of angular momentum and mass,
respectively. Assumed parameter values are listed in Table~\ref{tab:assumed_params}.

This wind model consists of only highly ionised material,
whose typical ionisation state is $\log\xi\sim5$, consistent with our 
implementation, 
where only H- and He-like ions of Fe and Ni are considered.
However, the observed spectra often show strong
continuum absorption at lower energies as shown in Fig.~\ref{fig:apm_pds}.
Such strong absorption requires much lower ionisation material, which are
not included in our wind model. For these, we additionally use a
partially ionised absorber, which partially covers the source.

\begin{table}
\caption{Assumed parameters for the simulations}
\centering
\begin{tabular}{lc}
\hline\hline
Parameter & Value\\
\hline
Acceleration index $\beta$ & 1.0\\
Turbulent velocity $v_{\rm t}$ & 1000~km~s$^{-1}$\\
Initial velocity $v_0$ ($=v_{\rm t}$) & 1000~km~s$^{-1}$\\
Covering fraction $\Omega/4\pi$ & 0.15\\
Minimum radius $R_{\rm min}$ & $\simeq 2/(v_\infty /c)^2R_{\rm g}$\\
Inner angle $\theta_{\rm min}$ & $45^\circ$\\
\hline
\end{tabular}
\label{tab:assumed_params}
\end{table}%

\subsection{Parameters for \apm} \label{sec:eddington}

Our wind model is self-similar in ionisation structure and column
density for systems at different mass but the same Eddington ratio, so 
$L_{\rm bol}/L_{\rm Edd}$ is the most important parameters for
our wind model \citep{Hagino2016}. 
However, the intrinsic Eddington ratio of \apm is not clear as there is a 
large uncertainty in the magnification factor $\mu$ from gravitational lensing.
Some papers report strong
magnification with $\mu\sim 100$ \citep{Egami2000,Weiss2007,Krips2007}, but others
claim much smaller values of $\mu\sim2$--10
\citep{Lewis2002,Solomon2005,Riechers2009}.
\cite{Saez2009} circumvented this uncertainty by using instead the
relation between the Eddington ratio, $L_{\rm bol}/L_{\rm Edd}$, and
X-ray photon index, $\Gamma$ \citep{Wang2004,Shemmer2006,Shemmer2008}.
According to this relation, they estimated the Eddington ratio to be
$L_{\rm bol}/L_{\rm Edd}\simeq0.2$--$0.3$ from the X-ray photon index of this source
$\Gamma \sim2.0$.

However, Fig.\ref{fig:apm_pds} shows that it is at least feasible that
the spectrum of \apm is affected by absorption up to $\sim
20$--$30$~keV so that $\Gamma$ and hence $L_{\rm bol}/L_{\rm Edd}$ are
underestimated.  We evaluate the intrinsic photon index by fitting the
5--8~keV spectra of all the observations. This energy range
corresponds to $\sim25$--40~keV in the restframe, where the continuum
spectrum should be mostly free from absorption.  The photon index is
tied across all the observations, but the normalization is allowed to be
free. All the normalizations are 
consistent with each other except for Epoch~4, where the flux is significantly higher
than the other observations.  This fit gives a photon index
$\Gamma=2.26_{-0.27}^{+0.28}$, corresponding to $L_{\rm bol}/L_{\rm
  Edd}\sim0.5$ based on the relation in \cite{Grupe2010}. 
This is a lower limit as any residual absorption means that the
intrinsic X-ray photon index is larger.
Therefore, the Eddington ratio in this source is
$L_{\rm bol}/L_{\rm Edd}\gtrsim0.5$, similar to that of \pds, so we use the same wind model
for \apm as for \pds \citep{Hagino2015}. This has parameters detailed in Table 2. 

\begin{figure}
\centering
\vspace{-1cm}
\includegraphics[width=1.1\hsize]{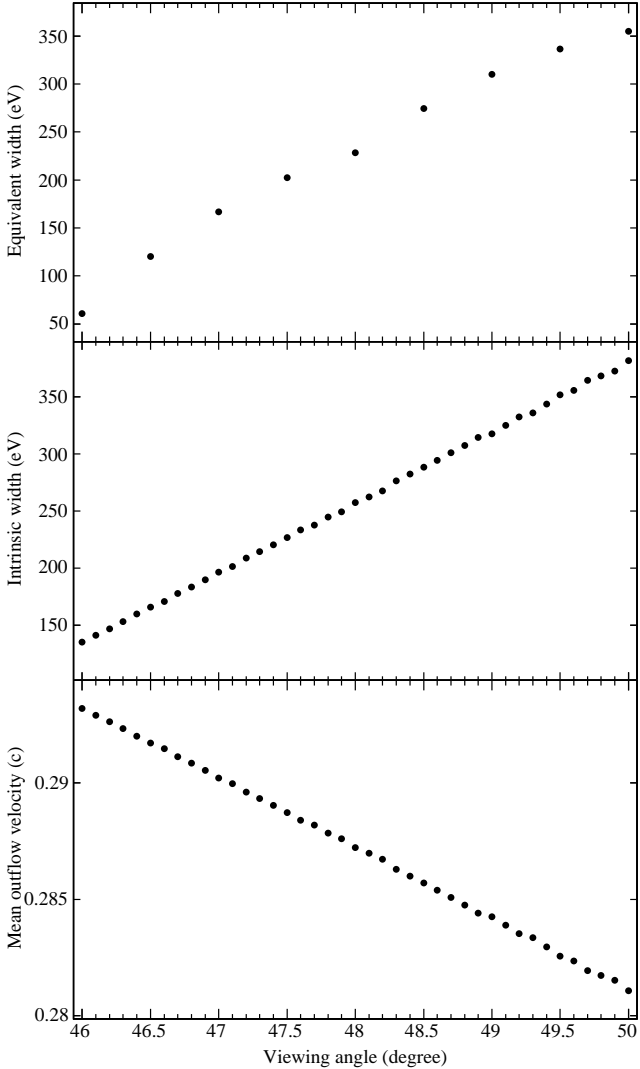}
\vspace{-0.5cm}
\caption{
Equivalent width (top panel), intrinsic width (middle panel) and mean outflow
velocity (bottom panel) of the blueshifted \fexxv/\fexxvi absorption lines as a function of
viewing angle of the wind. The intrinsic width is defined as the Full Width at
Zero Intensity (FWZI), which corresponds to the energy difference between
the fastest and slowest components of material along
the line of sight. The equivalent widths of \fexxv and \fexxvi are coadded
in this plot.
}
\label{fig:angle}
\end{figure}

The change in depth of the absorption lines in \pds and \hh from the hot wind can be 
reproduced by a changing viewing angle $\theta_{\rm incl}$
\citep{Hagino2015,Hagino2016}. As described
in detail in \cite{Hagino2016}, the energy of the absorption line depends on both the
terminal wind speed and the viewing angle, whereas the width of the absorption line
depends on the spread of velocities along the line of sight. Along the top edge of the
bicone, the line width is fairly small and the blueshift indicates the true wind velocity
since most of the wind is at its
terminal velocity. On the other hand, at higher viewing angles, the absorption line
is wider and the total blueshift is not so large since the line of sight cuts across
the acceleration region, where the velocity is much lower. Thus, changing only the
viewing angle gives very different observational properties of the absorption line
for the same wind model.
Fig.~\ref{fig:angle} shows quantitative results for the wind model used here in terms of
the iron K$\alpha$ line equivalent width, intrinsic width and velocity shift as a function of 
inclination angle. The
equivalent width is the sum of both H and He-like K$\alpha$ lines, as these merge
together for inclinations greater than 48$^\circ$, whereas the intrinsic width and
velocity shift are calculated for a single line. 

In this work, we use the spectral model created for \pds because its Eddington ratio
is similar to \apm. It is simulated for a wind with a mass outflow rate of
$\dot{M}_{\rm wind}/\dot{M}_{\rm Edd}=0.13$, a wind terminal velocity of $0.3c$, 
an ionising photon spectrum with $L_{\textrm{2--10~keV}}
/L_{\rm Edd}=1.6\times10^{-3}$ and $\Gamma=2.5$.
This is implemented as a multiplicative model in {\sc XSPEC} so it can approximately
describe the effect of disc wind on any similar continuum, and we similarly incorporate
any small change in velocity with a free redshift factor.

\section{Comparison of the Monte Carlo simulations and the observed spectra}
\subsection{Comparison with the Epoch 1 spectrum}

We first do a detailed fit to the spectrum of Epoch 1 as this has the
highest signal-to-noise absorption line detection.  Similarly to \pds
and \hh, the continuum absorption is modelled with absorption from
partially ionised material which partially covers the source ({\sc
  zxipcf}). This partially ionised absorber is required to reproduce
the strong continuum absorption at low energies in the observed
spectra which cannot be reproduced by our hot wind model.  We
additionally include the {\sc cabs} model because the high energy
continuum should be suppressed by Compton scattering which is not
included in {\sc zxipcf}. We assume that {\sc cabs} also partially
covers the source with a same covering factor as {\sc zxipcf}.

\begin{table*}
\caption{Fitting parameters for Epoch1}
\begin{center}
\begin{tabular}{llccc}
\hline\hline
 &  & {\sc zxipcf*wind*powerlaw} & {\sc zphabs*wind*powerlaw} & {\sc zphabs*zxipcf*wind*powerlaw}                                                           \\
\hline
Cold absorber$^a$ & $N_{\rm H}$ ($10^{22}$~cm$^{-2}$) & ---  & $6.2^{+0.8}_{-0.7}$ & $6.5^{+1.4}_{-3.0}$\\
Cool clump &  $v_{\rm out}$$^b$ ($c$) & $-0.09^{+0.32}_{-0.15}$ & --- & $0.19^{+0.07}_{-0.16}$\\
& $N_{\rm H}$ ($10^{22}$~cm$^{-2}$) & $2.9^{+8.1}_{-0.8}$ & --- & $42^{+67}_{-41}$\\
 & $\log\xi$ & $<0.1$ & --- & $1.6^{+2.8}_{-1.3}$\\　
 & $f_{\rm cov}$ & $>0.91$ & --- & $0.51^{+0.22}_{-0.16}$\\
Hot wind & $v_{\rm out}$ ($c$) & $0.17^{+0.01}_{-0.01}$ & $0.17^{+0.02}_{-0.02}$ & $0.17^{+0.01}_{-0.02}$\\
 & $\theta_{\rm incl}$ ($^\circ$) & $48.6^{+1.3}_{-1.0}$ & $47.8^{+1.1}_{-1.0}$ & $48.2^{+1.3}_{-1.8}$\\
Powerlaw & $\Gamma$ & $1.64^{+0.07}_{-0.06}$ & $1.63^{+0.06}_{-0.06}$ & $1.92^{+0.22}_{-0.29}$\\
 & Norm. ($10^{-4}$) & $1.0^{+0.1}_{-0.1}$ & $0.95^{+0.08}_{-0.07}$ & $1.8^{+1.7}_{-0.8}$\\
\hline
Fit statistics  & $\chi^2_\nu$   & 79.73 / 101 & 79.42 / 104 & 76.00 / 100\\
                   & Null Prob.            & 0.94 & 0.97 & 0.96\\
\hline
\end{tabular}
\begin{flushleft}
\footnotesize
$^a$ Model for the cool clumps consists of {\sc zxipcf} and {\sc cabs}.\\
$^b$ Minus sign means the inflow/redshift.
\end{flushleft}
\label{tab:Epoch1}
\end{center}
\end{table*}

\begin{figure}
\begin{center}
\includegraphics[width=1.0\hsize]{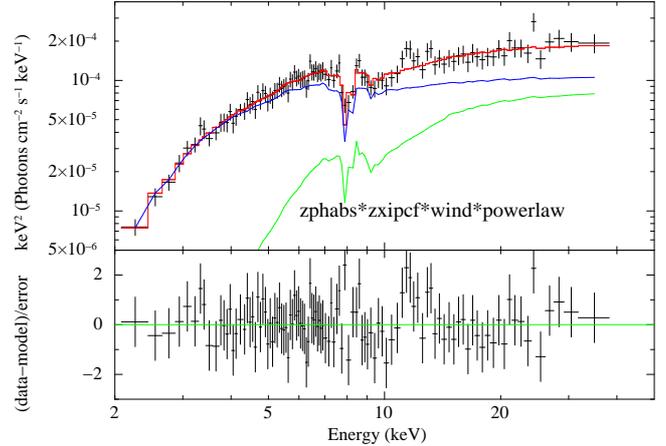}
\caption{Observed spectrum and best-fit model for Epoch~1. The absorbed
component, unabsorbed component and sum of these components are plotted
in green, blue and red, respectively.}
\label{fig:Epoch1}
\end{center}
\end{figure}

The partially ionised absorber is less ionised ($\log\xi<0.1$) and 
covers more of the source ($f>0.91$) than in \pds and \hh.
Full parameters are listed in the left column of
Table~\ref{tab:Epoch1}. The outflow velocity of the absorber is 
poorly constrained but is consistent
with zero ($v=-0.24c\textrm{--}0.23c$). We assume it is at rest, and use
the full covering neutral absorber {\sc zphabs} instead of {\sc zxipcf}.
This model provides a similar $\chi^2$ but has three fewer parameters
as listed in the center column of Table~\ref{tab:Epoch1}.
In both models, the outflow velocity of the hot wind is similar to the local fast wind
sources \citep{Tombesi2010,Gofford2013}, $0.18\pm0.02c$, and photon index of the
intrinsic powerlaw is very hard value of $\sim1.6$. Thus the sharp downturn
below 4~keV (which is not present in \pds or \hh, see Fig. 1) could be from an additional 
neutral screen of material. Hence we  
add again a partially ionised absorber in order to investigate the effect
of cool clumps in the wind. These were required in our previous studies and are also
naturally expected to exist in hot winds due to the ionisation instability
\citep{Krolik1981} or/and the Rayleigh-Taylor instability \citep{Takeuchi2014}.
The best fit parameters are listed in the last column of Table~\ref{tab:Epoch1},
and the observed spectrum and model are shown in Figure~\ref{fig:Epoch1}.
Although adding cool clumps does not improve the fit significantly
(F-test probability $\sim30\%$), this gives an interesting result. The outflowing
velocity of the partially ionised absorber is consistent with the hot wind velocity.
A similar result is found in \pds \citep{Matzeu2016}, and it is consistent with
our interpretation that the partially ionised absorber is due to cool clumps
embedded in the hot phase of the wind. In this model, the absorption line at $\sim8$~keV 
is mainly set by the hot wind, the feature at $\sim9$~keV and the 4--6~keV continuum shape are 
mainly set by the partially ionised absorption, while the continuum below 4~keV is 
set by the neutral absorption. 
Adding the partially ionised absorber does not change the parameters of
the neutral absorber or the hot wind, but the powerlaw continuum becomes
steeper.

\subsection{Application to all \xmm, \chandra and \suzaku data}
\begin{table*}
\caption{Fitting parameters for all the \xmm and \chandra observations}
\begin{center}
\begin{tabular}{llcccccc}
    \hline\hline
     &                                     & Epoch 0                                                           & Epoch 1                        & Epoch 2                         & Epoch 3                         & Epoch 4                         & Epoch 5                         \\
     \hline
Cold absorber & $N_{\rm H}$ ($10^{22}$~cm$^{-2}$) & $5.9^{+0.8}_{-1.5}$ & $6.4^{+1.4}_{-4.6}$ & $5.4^{+0.9}_{-0.9}$ & $5.5^{+0.7}_{-0.7}$ & $5.2^{+0.5}_{-0.5}$ & $5.6^{+1.2}_{-1.1}$\\
Cool clump &  $v_{\rm out}$ ($c$) & \multicolumn{6}{c}{tied to hot wind}\\
 & $N_{\rm H}$ $(10^{22}$~cm$^{-2}$) & $87^{+208}_{-67}$ & $44^{+60}_{-42}$ & $87^{+43}_{-38}$ & $>254$ & $138^{+163}_{-42}$ & $76^{+159}_{-47}$\\
 & $\log\xi$ & $<2.9$ & $1.9^{+2.0}_{-2.0}$ & $2.3^{+0.7}_{-0.3}$ & $2.8^{+0.2}_{-1.4}$ & $2.9^{+0.3}_{-0.6}$ & $<2.2$\\
 & $f_{\rm cov}$ & $0.52^{+0.24}_{-0.26}$ & $0.46^{+0.25}_{-0.23}$ & $0.58^{+0.14}_{-0.22}$ & $0.93^{+0.04}_{-0.08}$ & $0.61^{+0.20}_{-0.16}$ & $0.44^{+0.24}_{-0.34}$\\
Hot wind & $v_{\rm out}$ ($c$) & $0.22^{+0.03}_{-0.03}$ & $0.17^{+0.02}_{-0.02}$ & $0.17^{+0.01}_{-0.01}$ & $0.11^{+0.02}_{-0.02}$ & $0.11^{+0.01}_{-0.01}$ & $0.10^{+0.05}_{-0.05}$\\
 & $\theta_{\rm incl}$ ($^\circ$) & $47.8^{+1.8}_{-1.4}$ & $48.1^{+1.3}_{-1.3}$ & $47.5^{+0.8}_{-0.7}$ & $<46.4$ & $46.8^{+0.5}_{-0.5}$ & $<46.9$\\
Powerlaw & $\Gamma$ & $2.11^{+0.33}_{-0.23}$ & $1.87^{+0.33}_{-0.26}$ & $2.11^{+0.18}_{-0.16}$ & $2.24^{+0.09}_{-0.09}$ & $2.22^{+0.09}_{-0.08}$ & $2.06^{+0.23}_{-0.22}$\\
 & Norm. ($10^{-4}$) & $3.0^{+2.7}_{-1.3}$ & $1.6^{+1.6}_{-0.7}$ & $2.3^{+1.2}_{-0.8}$ & $19^{+31}_{-11}$ & $4.5^{+5.3}_{-1.2}$ & $2.2^{+1.7}_{-0.9}$\\
                   \hline
    Fit statistics  & $\chi^2_\nu$   & 79.99 / 58 & 76.06 / 101 & 133.33 / 148 & 124.24 / 141 & 172.42 / 177 & 101.09 / 112\\
                   & Null Prob.                           & 0.029 & 0.97 & 0.80 & 0.84 & 0.58 & 0.76\\
                     & $\chi^2/\nu$   & 1.38 & 0.75 & 0.90 & 0.88 & 0.97 & 0.90\\
\hline
\cite{Saez2011} & $\chi^2/\nu$ (extreme wind) & ---  & 1.15 & 0.95 & 1.03 & 1.08 & 0.97\\
                    \hline
    \end{tabular}
    \end{center}
\label{tab:allEpoch}
\end{table*}

\begin{table*}
\caption{Fitting parameters for all the \suzaku observations}
\begin{center}
\begin{tabular}{llccc}
\hline\hline
     &                                     & OBS1 & OBS2$^a$ & OBS3\\
\hline
Cold absorber & $N_{\rm H}$ ($10^{22}$~cm$^{-2}$) & $7.4^{+1.9}_{-1.4}$ & $5.6^{+1.6}_{-1.9}$ & $5.3^{+1.5}_{-1.7}$\\
Cool clump &  $v_{\rm out}$ ($c$) & \multicolumn{3}{c}{tied to hot wind}\\
 & $N_{\rm H}$ ($10^{22}$~cm$^{-2}$) & $191^{+89}_{-79}$ & $92^{+54}_{-40}$ & $102^{+258}_{-56}$\\
 & $\log\xi$ & $<3.0$ & $<2.4$ & $1.9^{+2.4}_{-2.3}$\\
 & $f_{\rm cov}$ & $0.81^{+0.11}_{-0.13}$ & $0.59^{+0.18}_{-0.36}$ & $0.57^{+0.08}_{-0.39}$\\
Hot wind & $v_{\rm out}$ ($c$) & $0.12^{+0.07}_{-0.06}$ & $0.18^{+0.05}_{-0.04}$ & $0.14^{+0.03}_{-0.03}$\\
 & $\theta_{\rm incl}$ ($^\circ$) & $<46.5$ & $<47.2$ & $<46.7$\\
Powerlaw & $\Gamma$ & $2.28^{+0.20}_{-0.16}$ & $2.17^{+0.25}_{-0.26}$ & $2.17^{+0.21}_{-0.31}$\\
 & Norm. ($10^{-4}$) & $6.8^{+8.9}_{-4.0}$ & $2.8^{+2.5}_{-1.5}$ & $2.7^{+1.7}_{-1.4}$\\
                   \hline
    Fit statistics  & $\chi^2_\nu$  & 125.02 / 130 & 130.60 / 124 & 161.75 / 142\\
                   & Null Prob.  & 0.61 & 0.32 & 0.12\\
                   & $\chi^2/\nu$  & 0.96 & 1.05 & 1.14\\
                   \hline
 \end{tabular}
 \begin{flushleft}
\footnotesize
$^a$ Data points of FI between 1.75--1.95~keV are ignored as in \cite{Saez2009}.
\end{flushleft}
\end{center}
\label{tab:allEpoch_suzaku}
\end{table*}

\begin{figure*}
\begin{center}
\includegraphics[width=0.32\hsize]{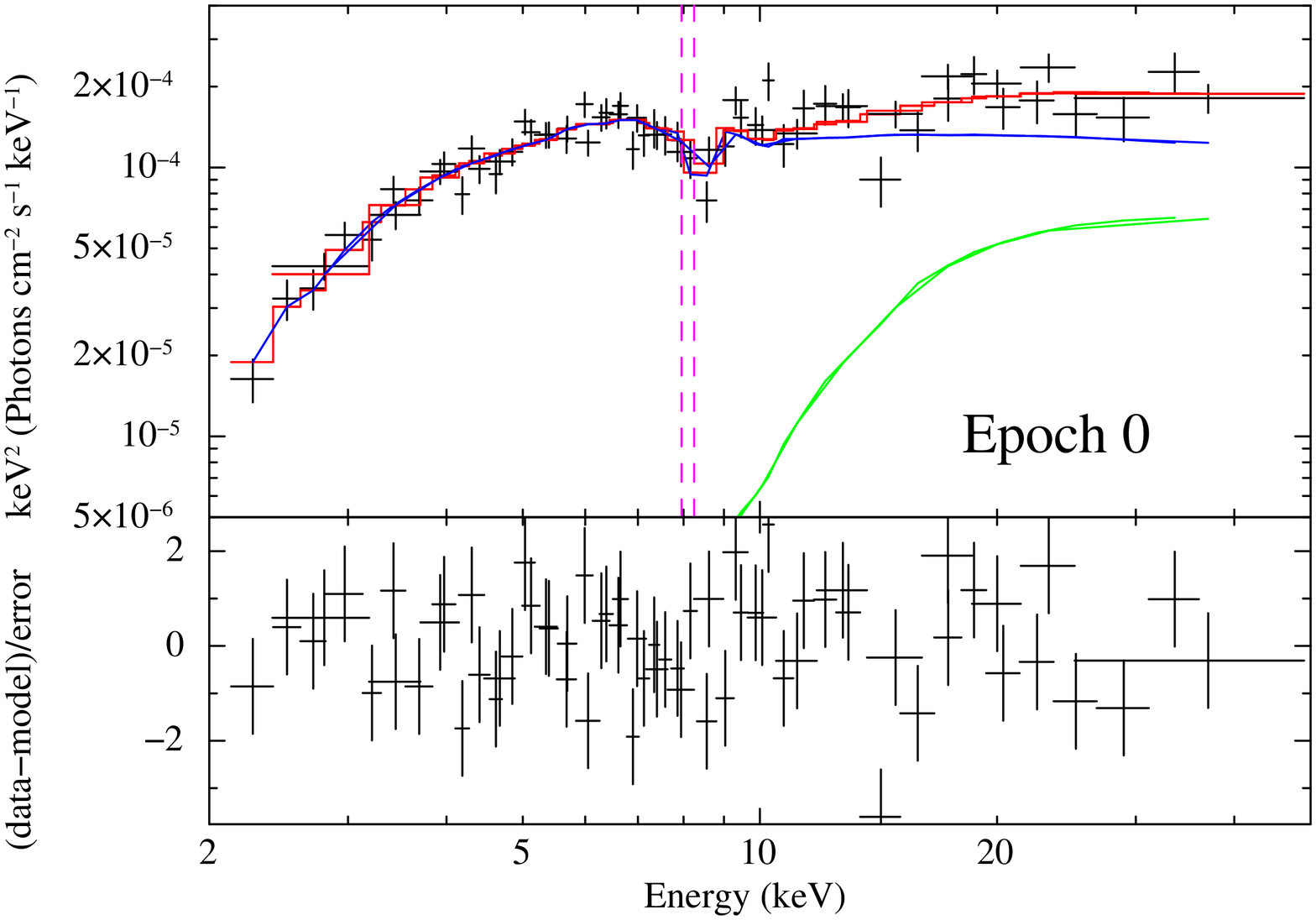}
\includegraphics[width=0.32\hsize]{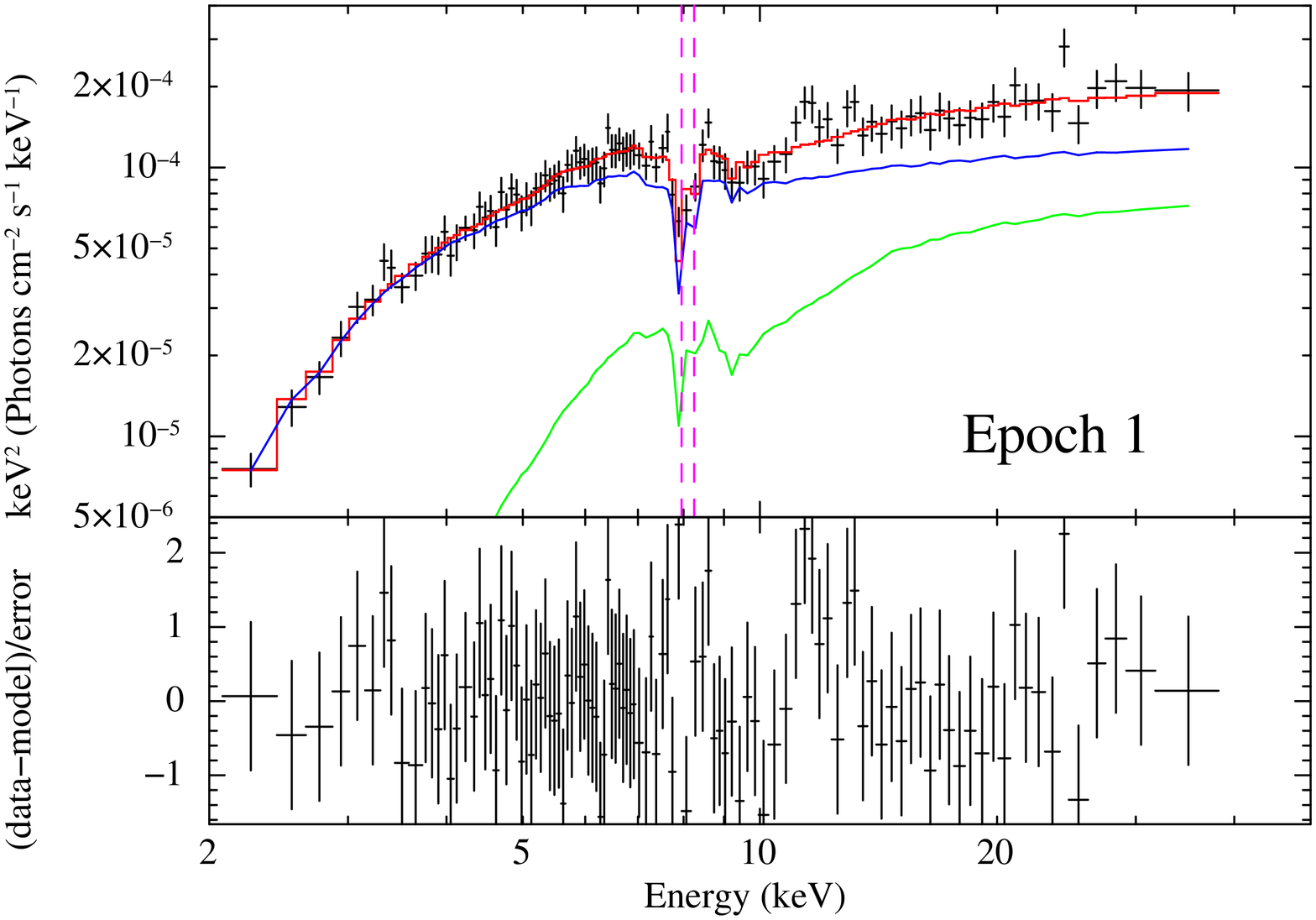}
\includegraphics[width=0.32\hsize]{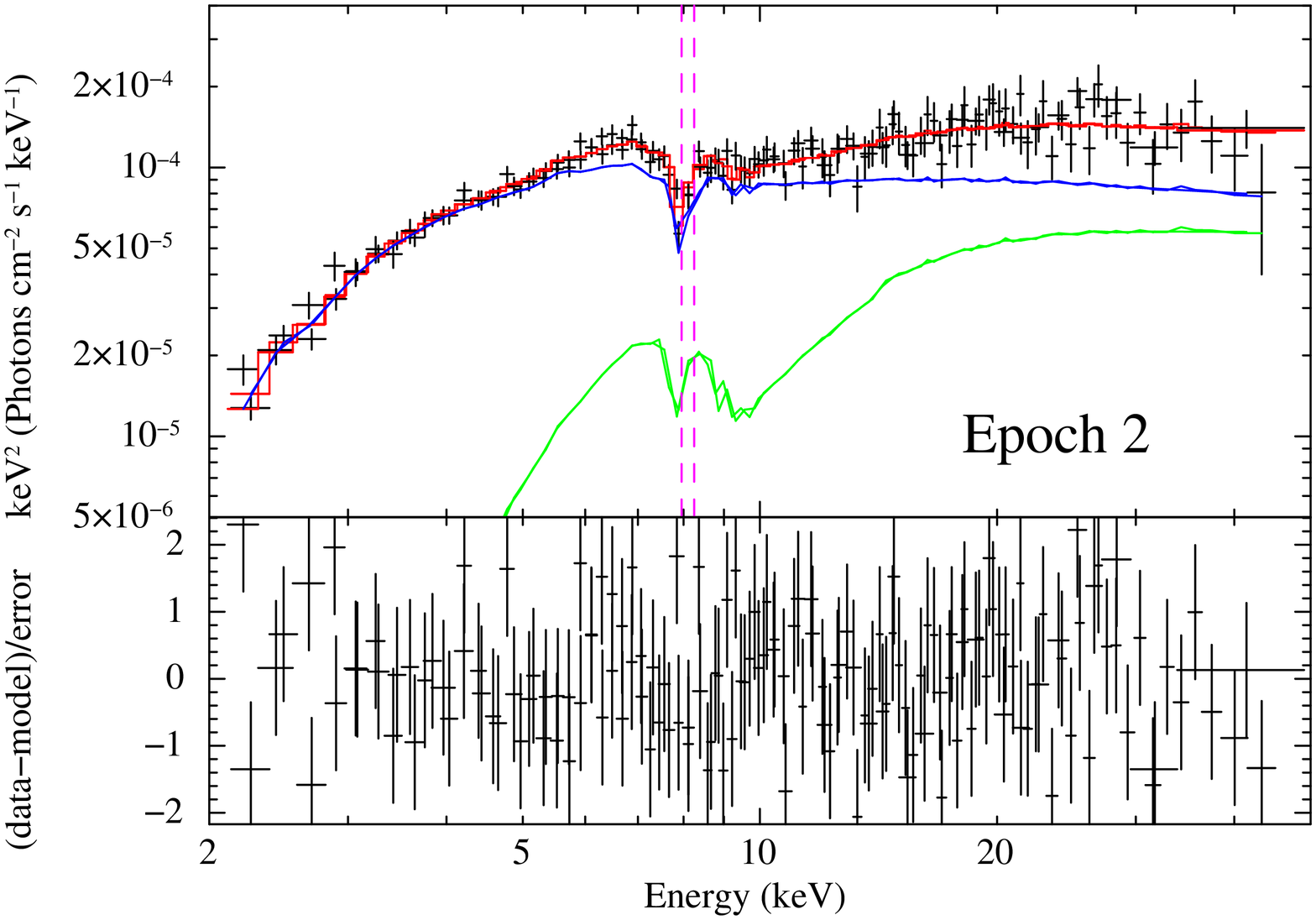}
\includegraphics[width=0.32\hsize]{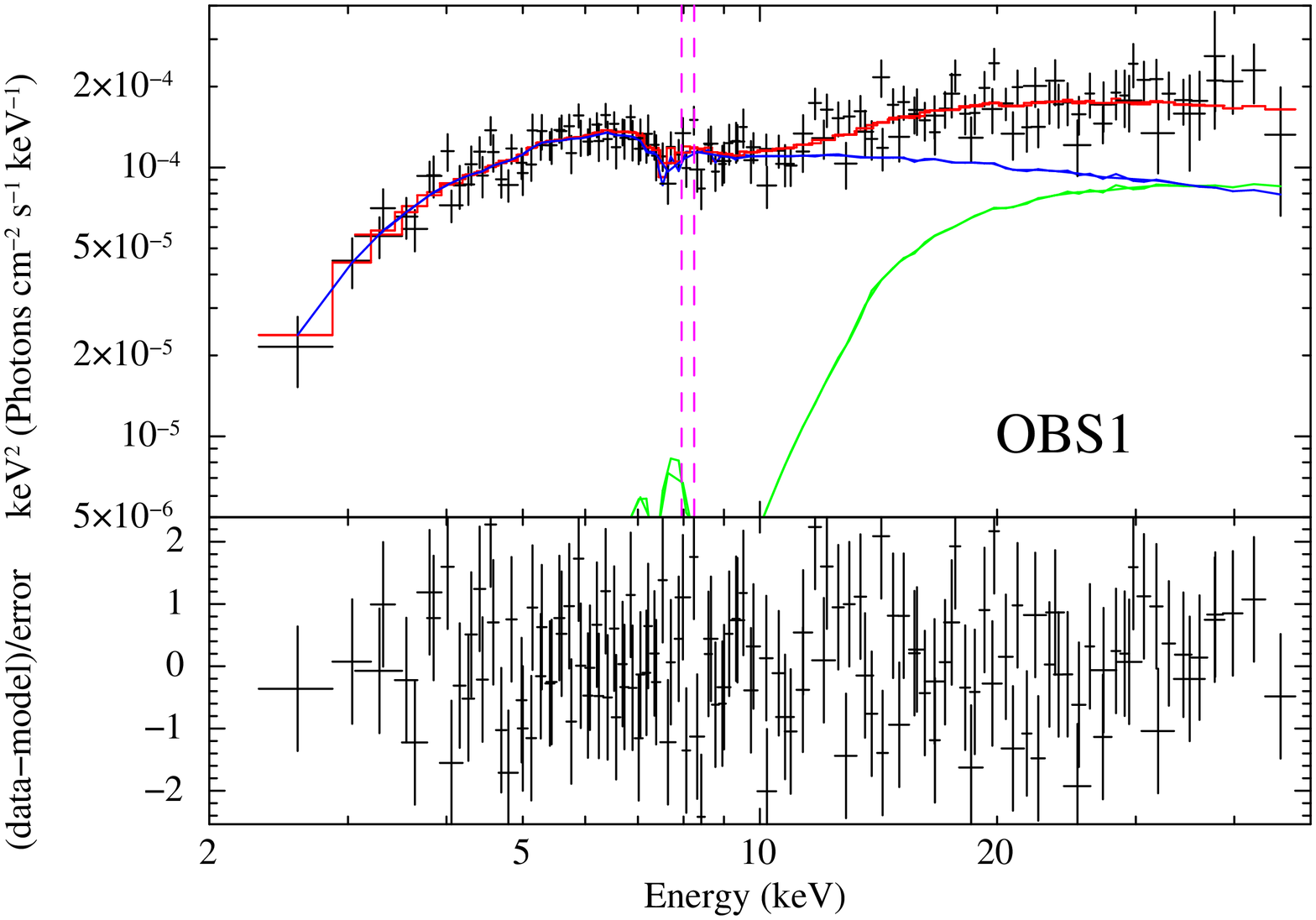}
\includegraphics[width=0.32\hsize]{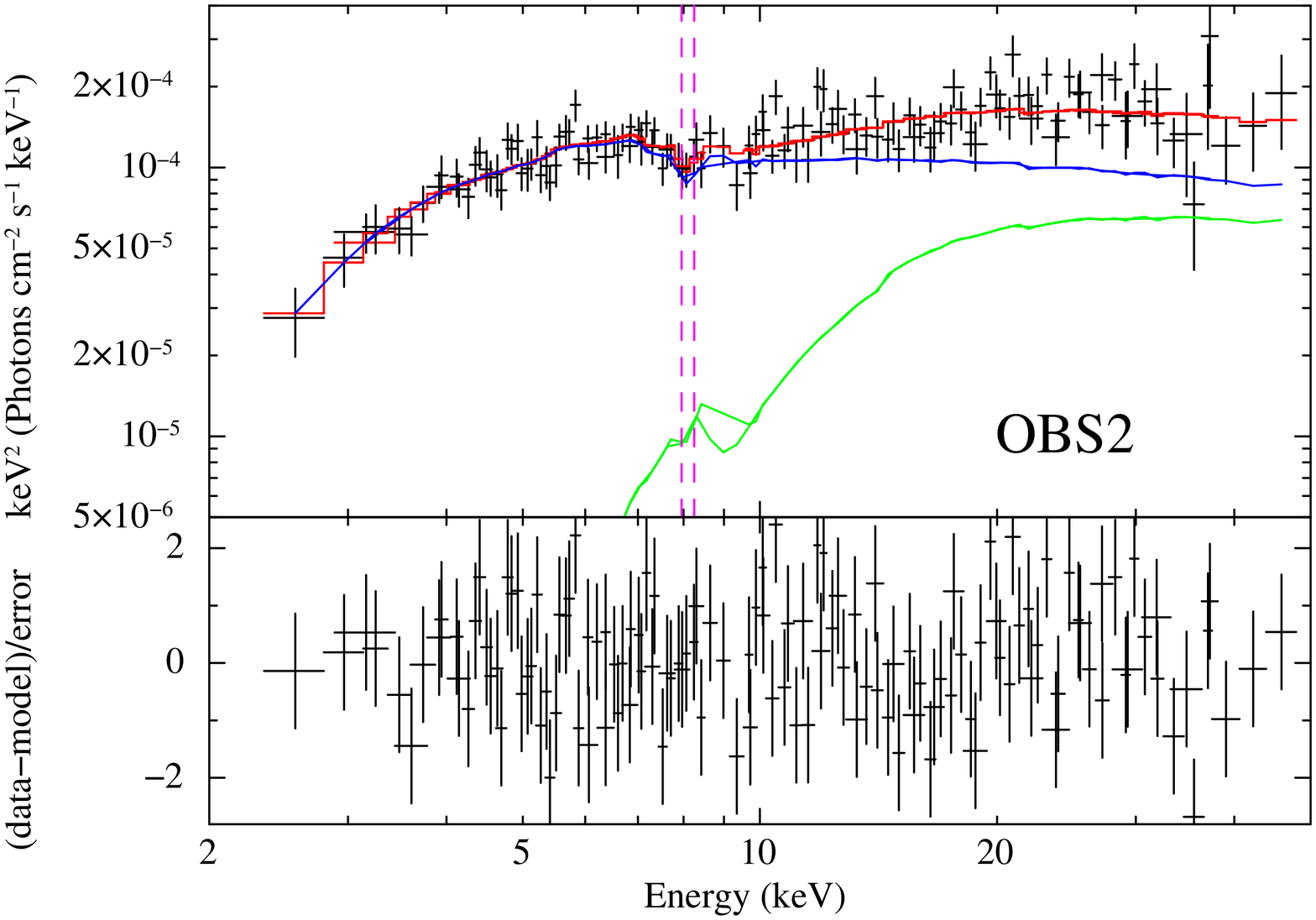}
\includegraphics[width=0.32\hsize]{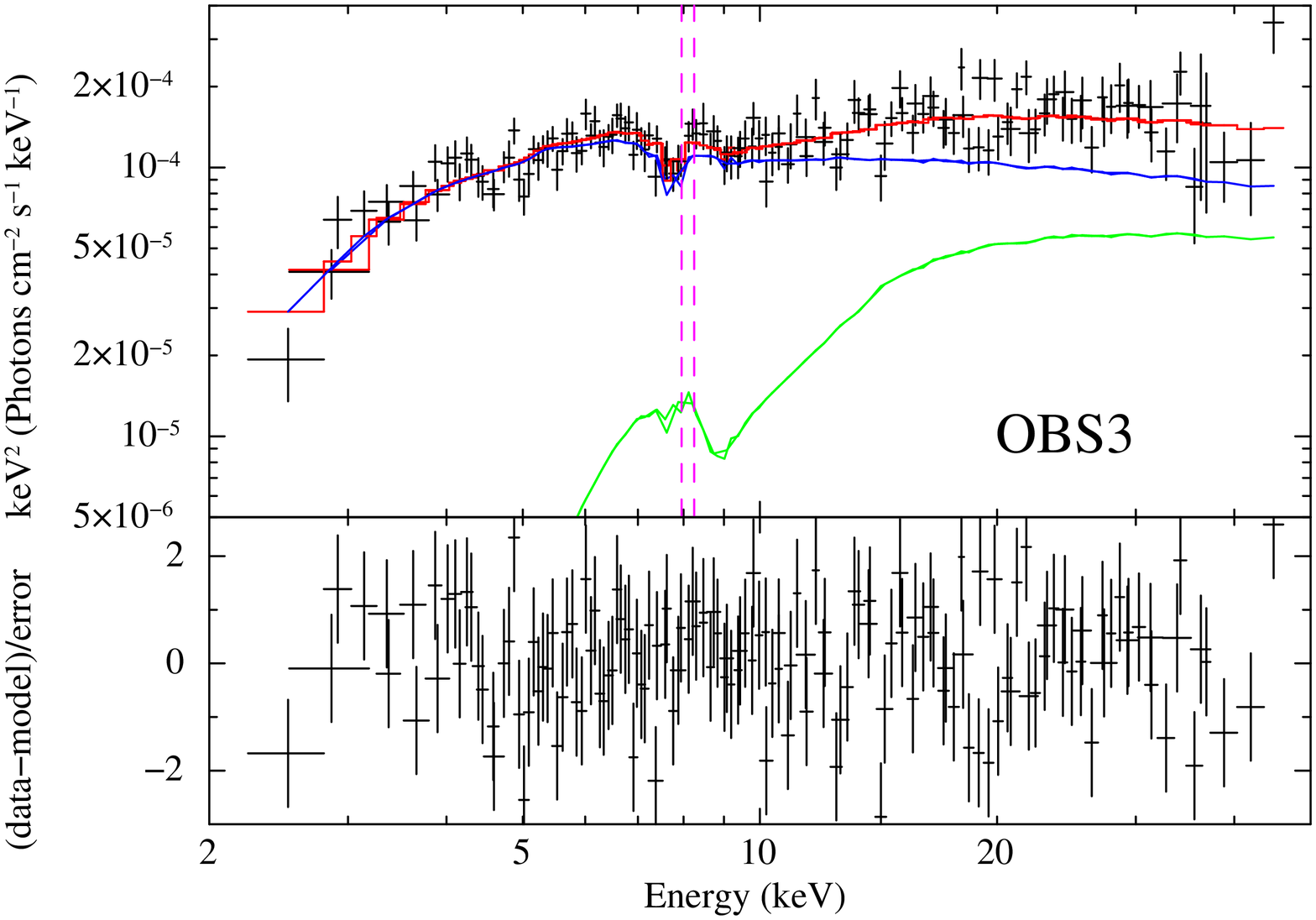}
\includegraphics[width=0.32\hsize]{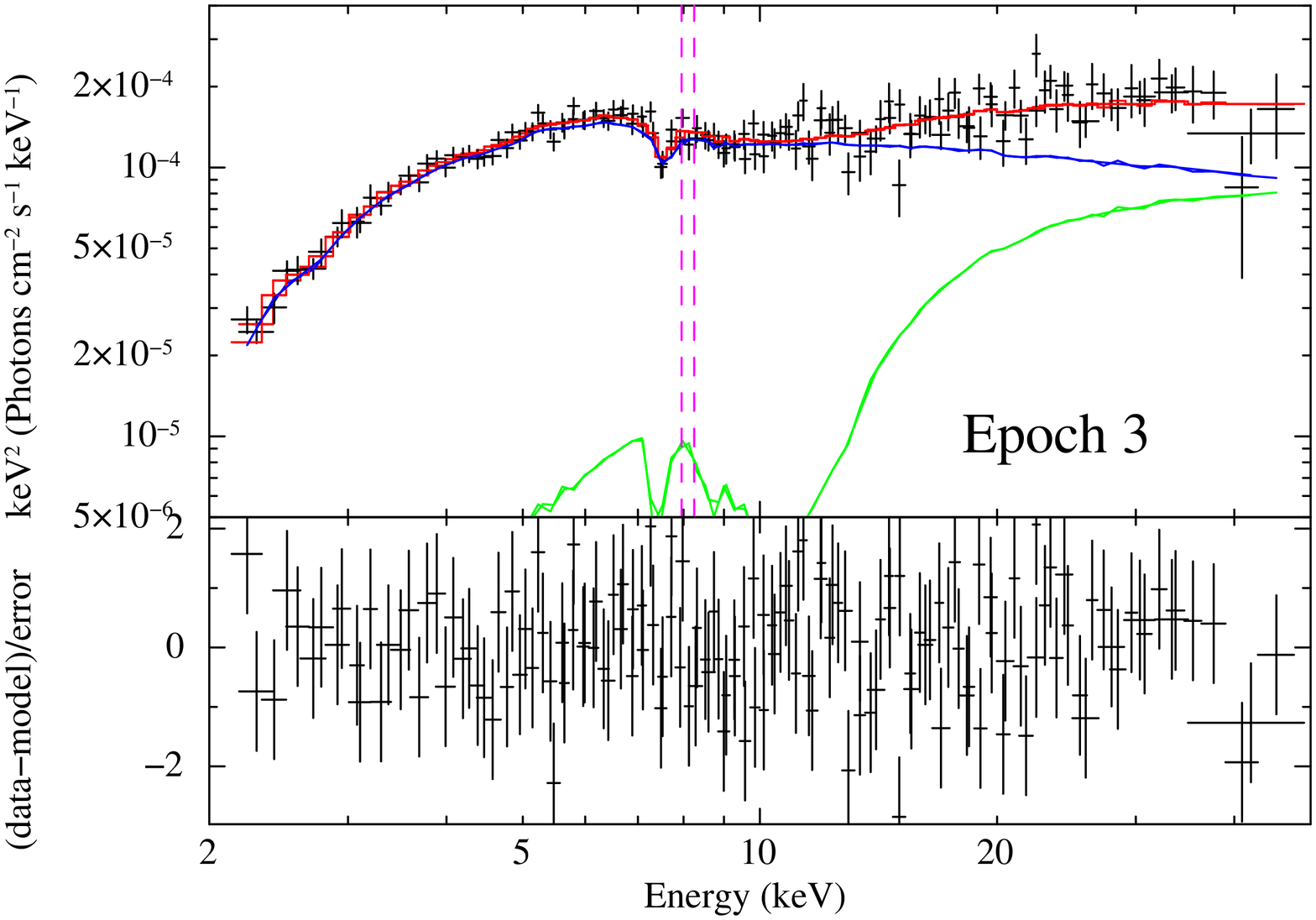}
\includegraphics[width=0.32\hsize]{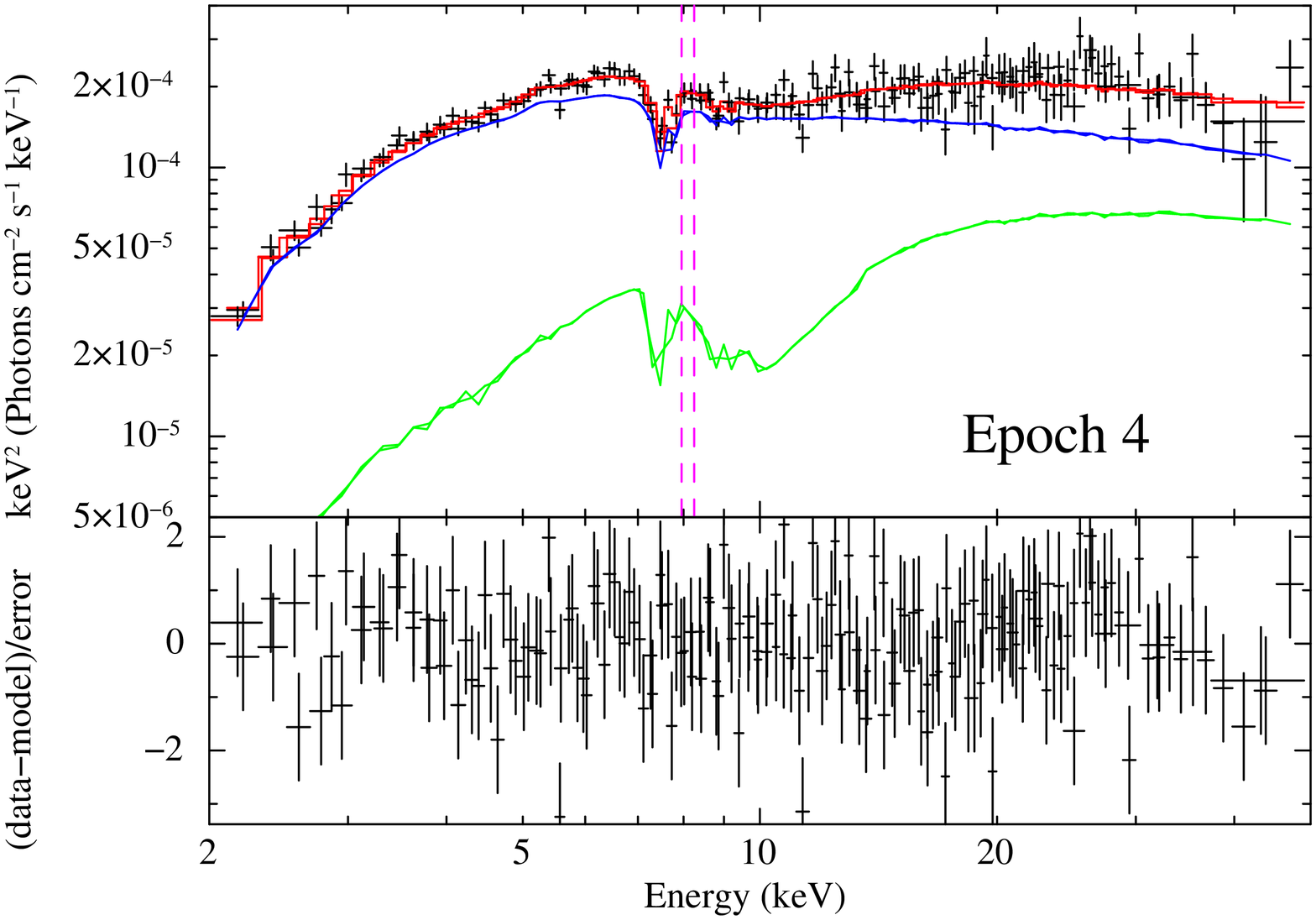}
\includegraphics[width=0.32\hsize]{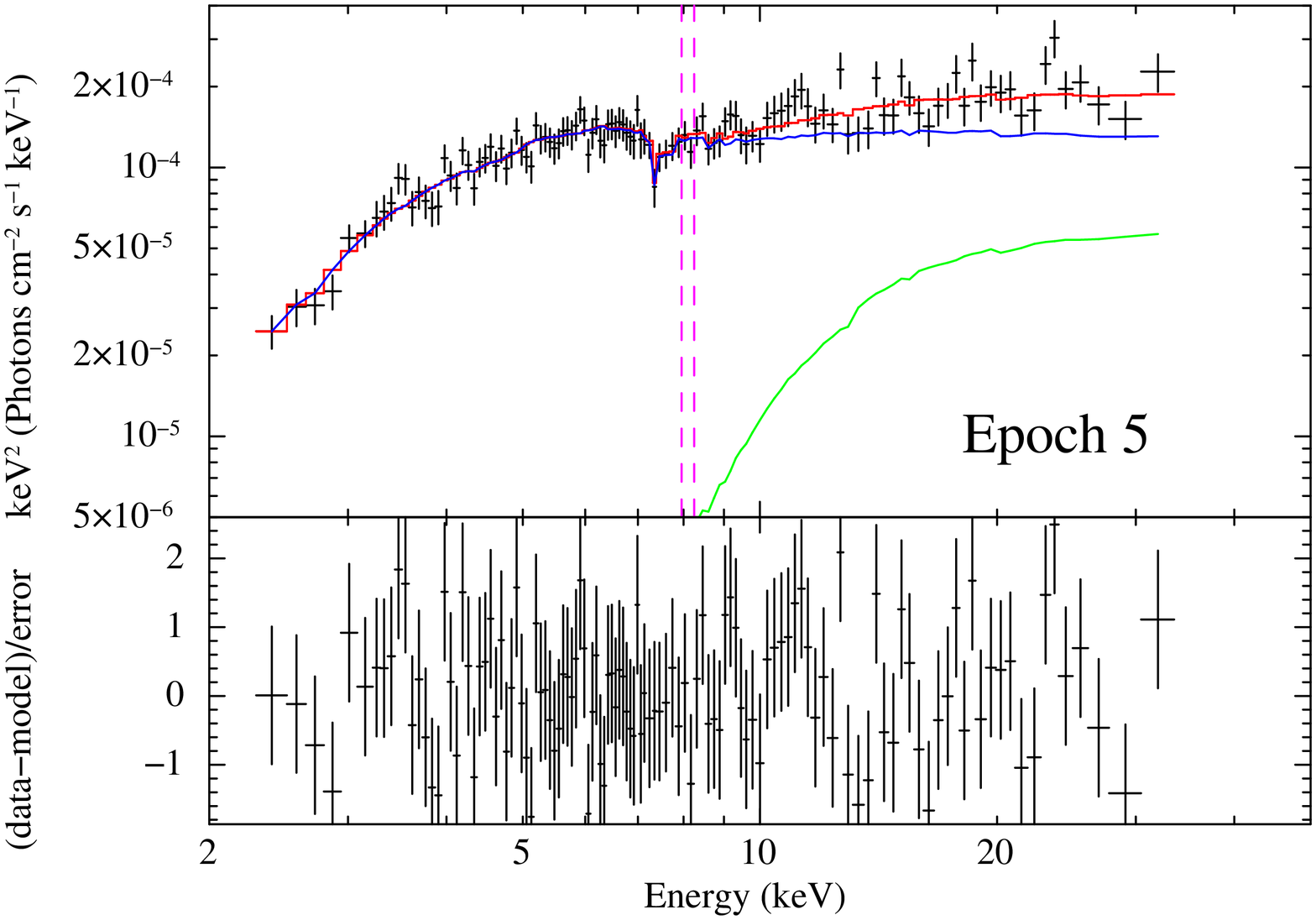}
\caption{Observed spectra and best-fit models for all Epochs. The absorbed
component, unabsorbed component and sum of these components are plotted
in green, blue and red, respectively. Magenta dashed lines indicate the best-fit
energies of the blueshifted \fexxv/\fexxvi absorption lines seen in Epoch~1. It is clear that the
velocity of these systems is decreasing over time, as is their equivalent width. }
\label{fig:allEpoch}
\end{center}
\end{figure*}

\begin{figure}
\begin{center}
\includegraphics[width=\hsize]{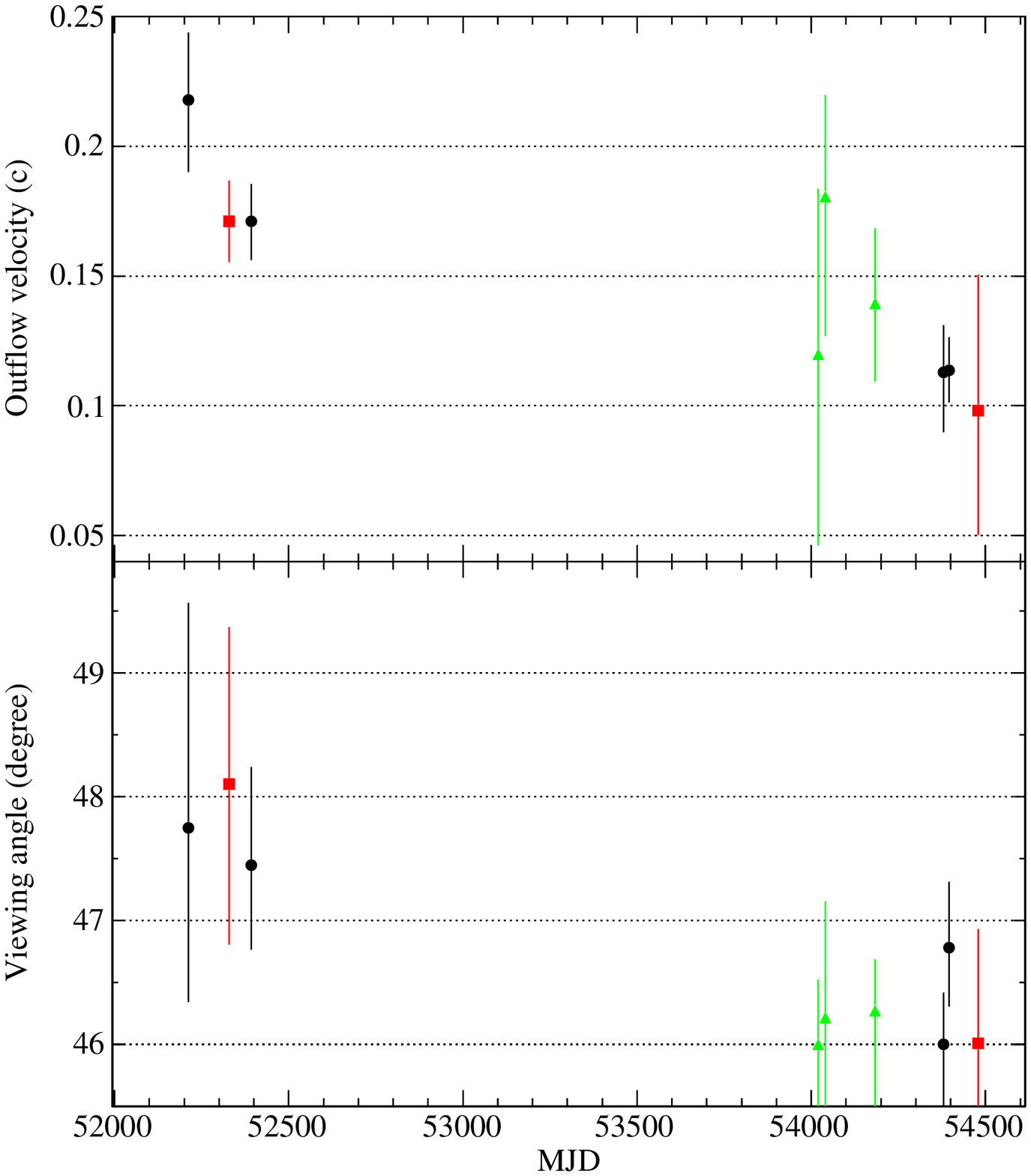}
\caption{
Outflow velocity and viewing angle of each observation 
(\xmm: black circles, \chandra: red squares, \suzaku: green triangles)
plotted as a function
of MJD. Error bars correspond to 90\% confidence level. 
}
\label{fig:deceleration}
\end{center}
\end{figure}

We use the model derived above for Epoch~1 to fit 
all the data observed by \xmm, \chandra and \suzaku.
The velocity of the partially ionised absorber is tied to that
of the hot wind since they are consistent in Epoch~1 spectrum.
The best-fit parameters for all the Epochs are listed in Table~\ref{tab:allEpoch} and
\ref{tab:allEpoch_suzaku}, and the spectra and models are shown in
Figure~\ref{fig:allEpoch}. Magenta dashed lines indicate the best-fit energies of the
hot wind absorption lines for Epoch~1, where the absorption lines are most clearly
detected. It is obvious  that the line energies of the hot wind decrease over 
time. In the observations in 2001 and 2002 (Epoch~0--2), the absorption lines are
the same or higher energies than those of Epoch~1, while in the observations after 2006
(OBS1--3, Epoch3--5) the line energies are lower than those of Epoch~1.
Also obviously, the depths of the absorption lines decrease in the later observations.

The column density of the cold absorber is fully consistent with
constant.  This is different from the previous study
\citep{Chartas2009} due to a significant improvement of the
contamination models of \chandra. The old {\sc CALDB} they used for
Epoch~5 underestimates effects of the contamination (G. Chartas,
private communication).  The intrinsic continuum is also mainly consistent
with being constant except for Epoch~3, which has
a much higher intrinsic powerlaw flux. 
We think that this is an artifact as the observed high energy
flux does not show such large variability as discussed in
Sec.~\ref{sec:eddington}. Instead, this is probably an artifact of our
approximate model for the cool clump absorption as the column density of the 
partial coverer has also increased dramatically. Electron
scattering from the clumps, which is currently modelled by {\sc cabs},
strongly depends on the geometry.  {\sc cabs} only considers photons
scattering out of our line of sight, but photons scattering into our
line of sight could be important if the solid angle of the wind is not
negligible.  We demonstrate this by refitting the spectrum in Epoch~3 without the
{\sc cabs} component. This gives a smaller powerlaw normalization of
$2.4^{+0.5}_{-0.4}\times10^{-4}$ and a smaller covering factor of
$0.44^{+0.07}_{-0.08}$, both of which are similar to the other
observations. The other parameters are consistent within 90\%
uncertainties with comparable fit statistics of
$\chi_\nu^2=124.0/141$.

The partially ionised absorber is moderately ionised ($\log\xi\sim2$)
in all the observations. It is less ionised than the hot wind, whose
ionisation parameter is typically $\log\xi\sim5$. It produces
absorption edges at energies lower than those of the H- and He-like
iron in the hot wind, and distorts the continuum spectral shape.  The
continuum shape is strongly affected by covering factor and column
density as well as the ionisation parameter. Due to this model
component, the intrinsic powerlaw continuum is steeper than in
\cite{Chartas2009}. On average, the photon index is
$\Gamma\sim2.2$, similar to that measured directly in the high energy
spectra (Sec.~\ref{sec:eddington}). This is 
slightly smaller than the $\Gamma=2.5$ assumed to derive the ionisation state in our 
disk wind model, but this only makes a  10--20\% difference in ionization parameter
$\log\xi$. This effect is much smaller than the factor 10 uncertainty on ionisation parameter which comes
from the uncertainty in intrinsic luminosity due to the lens magnification. 

The hot wind velocity clearly decreases from $\sim0.2c$ to $\sim0.1c$ during all the
observation Epochs, as shown in the top panel of Figure~\ref{fig:deceleration}. 
This is not an artifact of the correlated change in inclination angle from $\sim48^\circ$
to $\sim46^\circ$ (lower panel of Fig.~\ref{fig:deceleration}) as this corresponds to
$\Delta v\simeq0.004c$, which is much smaller than the decrease in the outflow
velocity (Fig.~\ref{fig:angle}).
The decreasing angle is instead a consequence of a decreasing column density of
the hot wind.

\section{Discussion}
\subsection{Velocity of the wind}
These data were previously fit by \cite{Chartas2009} and \cite{Saez2011}.
Our results agree fairly well in terms of the velocity of what we call the hot wind
component and they call the slow wind (the component which produces the obvious
absorption line in most of the spectra). However, they differ dramatically on how to
interpret the rest of the complex absorption at higher energies. In our model, there
is additional curvature from the edge structure from a partial covering, less ionised
component, which we assume is outflowing at the same velocity as 
our hot wind as seen in the classic wind source \pds \citep{Matzeu2016}. 
Instead, in \cite{Saez2011}, this broad absorption feature is again fit by
an iron resonance absorption line, so the observed width of the absorption requires
a large range of velocities in the line of sight in this second wind component. The
fastest material typically reaches speeds of 0.65--0.7$c$ except in Epoch~1, where
they only require $0.4c$ (their fast wind component). Thus in their model there is
material which is typically much faster than can be explained by any radiatively
driven wind, whereas in ours this is not required. 

The key question is then which wind model better matches the physical situation in
this source. All models are only approximations to a more complex reality, but it is
clearly useful to ask which one describes the data better. Our data have different
numbers of points due to differences in extraction and grouping, so we include
the reduced $\chi^2/\nu$ for the \cite{Saez2011} model fits for Epoch~1--5
at the bottom of Table~\ref{tab:allEpoch}. Our fits all
have lower $\chi^2_\nu$, despite there only being 5 free parameters to describe
our complex wind (velocity, angle which controls the column density and velocity
width of the hot wind, and then the cooler wind column, ionisation state and
covering fraction), compared to 7 in their model (each wind has column density,
and minimum and maximum outflow velocity, and then both winds are assumed
to have the same ionisation state). 

The data clearly show that there are two absorption features. All models agree
that the lower energy feature is mainly a resonance absorption iron line from material 
outflowing
at 0.1--0.2$c$. This has enough energy to impact the host galaxy and is clear evidence for
AGN feedback. The higher energy feature is more controversial. In our model,
it is produced by the complex edge feature from the less ionised, partial covering
material outflowing at the same velocity. These velocities are high but can be
produced by radiation driving on a wind launched from inner disc. In the
\cite{Saez2011} model, it is instead produced by a highly broadened absorption
line which requires extreme velocities. Not only does our model gives better fit,
but we note that the classic wind source, \pds requires such partially ionised
material which partially covers the source \citep{Reeves2009,Hagino2015} and
which is outflowing along with the material producing the resonance line
\citep{Matzeu2016}. This supports our interpretation, but we need data from \apm
which is of similar quality to
that of \pds in order to unambiguously distinguish between our model and the
extreme wind.

\subsection{Broadband SED and quasar parameters}
The broadband spectral energy distribution (SED) of the quasar is very
important to understand the acceleration mechanisms of the
wind. Strong UV radiation easily launches a wind by radiation pressure
on UV line transitions, while strong X-rays suppress it. Hence an SED
which is UV bright and X-ray weak is clearly consistent with UV line
driving, while one which has strong X-ray flux is less favourable.

\begin{figure}
\begin{center}
\includegraphics[width=\hsize]{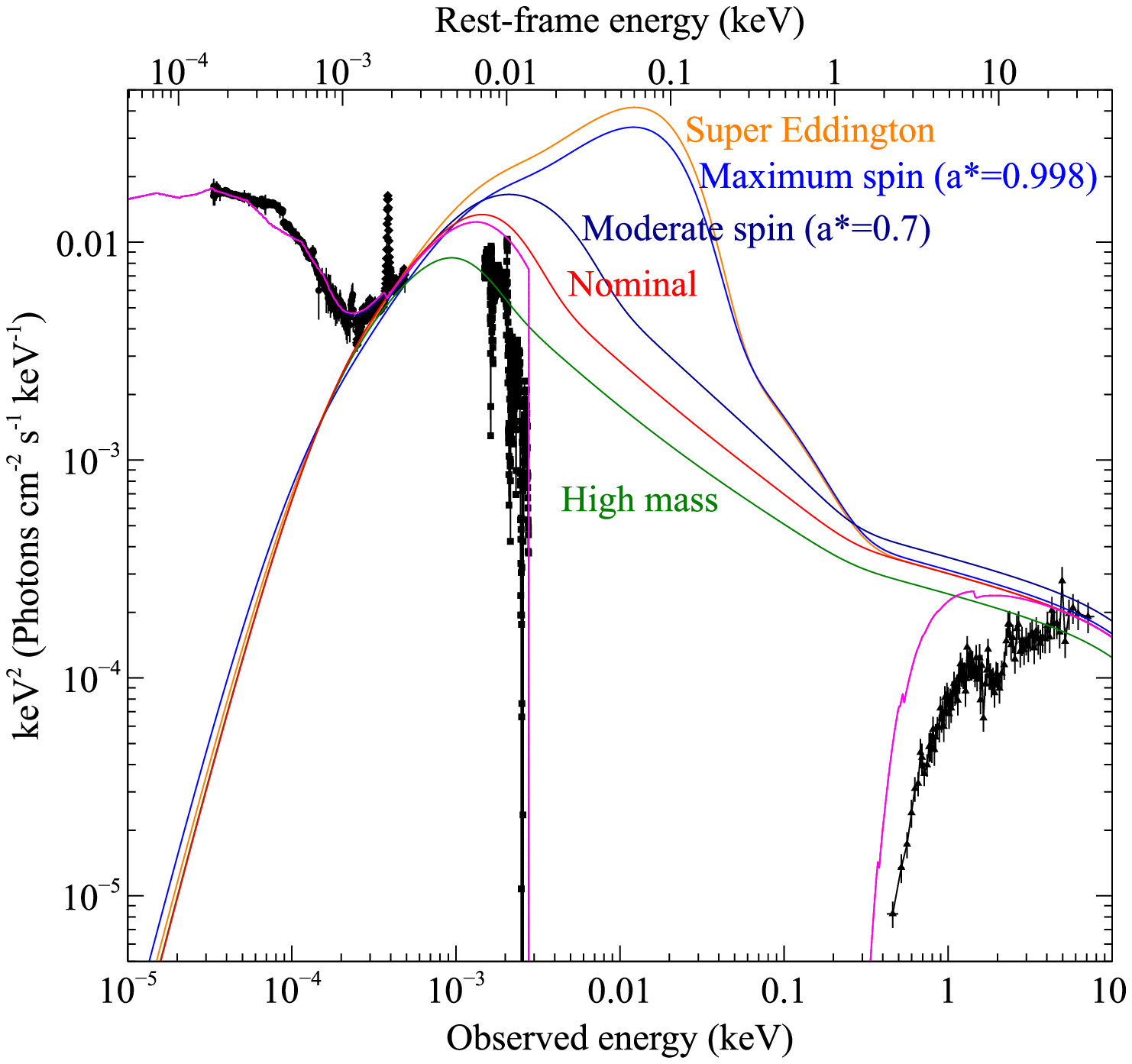}
\caption{Broadband spectral energy distribution of \apm from {\it
    Spitzer} \citep[black circles:][]{Soifer2004}, {\it AKARI}
  \citep[diamonds:][]{Oyabu2009}, {\it INT} \citep[squares:][]{Benn2002}
  and \chandra \citep[triangles:][]{Chartas2002}. The red line is model
  for the accretion flow with nominal parameters ($M_{\rm
    BH}=10^{10}$\msun, $L_{\rm bol}/L_{\rm Edd}=1$, $\mu=6$), while
  the magenta line shows reddening/absorption from our Galaxy and
  absorption by the constant gas column of $5\times 10^{22}$~cm$^{-2}$
  seen in the X-ray data. We also include template model to fit the IR
  torus emission in order to determine its luminosity.  The green line
  shows the upper limit to black hole mass for zero spin
  of $M_{\rm BH}=2\times10^{10}$\msun
  ($L_{\rm bol}/L_{\rm Edd}=0.5$, $\mu=4$), as the
  optical/UV continuum has to be at least as strong as the observed
  {\it INT} flux (it can be higher as there is substantial Lyman alpha
  forest absorption). Both Nominal and high mass models underpredict the 
 observed IR flux, requiring that the SED peaks at higher
  energies.  The dark blue line shows a solution with moderate black
  hole spin $a^*=0.7$ ($M_{\rm BH}=10^{10}$\msun, $L_{\rm bol}/L_{\rm
    Edd}=1$ and $\mu=8$), while the blue line shows maximum black hole
  spin $a^*=0.998$ ($M_{\rm BH}=10^{10}$\msun, $L_{\rm bol}/L_{\rm
    Edd}=1$ and $\mu=15$).  The orange line shows instead the lower
  limit of black hole mass $M_{\rm BH}=4\times10^{9}$\msun and high
  Eddington ratio of $L_{\rm bol}/L_{\rm Edd}=25$ ($a^*=0$ and
  $\mu=2$). Full parameters for these are listed in
   Tab.~\ref{tab:sedmodel}.
}
\label{fig:broadband}
\end{center}
\end{figure}

We plot the broadband SED from {\it Spitzer} \citep{Soifer2004}, {\it
  AKARI} \citep{Oyabu2009}, {\it 2.5~m Isaac Newton Telescope} ({\it
  INT}, \citealt{Benn2002}) and \chandra (Epoch1) data. The optical
continuum underneath the H$\alpha$ line shows a clear disc spectrum,
so we use this to constrain the mass accretion rate through the outer
accretion disc via the {\sc optxagnf} model \citep{Done2012}. This
includes phenomenological modelling of the soft X-ray excess and high
energy corona emission, assuming that these are energetically powered
by the same mass accretion rate as required for the outer disc. This
implies a transition radius, $R_{cor}$, within which the energy
released by gravity is dissipated in these X-ray components rather
than the standard disc. We first fix the black hole mass to $M_{\rm
  BH}=1.00^{+0.17}_{-0.13}\times10^{10}$\msun and a magnification
factor of $\mu\lesssim8$ as obtained by reverberation mapping of the
Si~\textsc{iv} and \civ emission lines \citep{Saturni2016}. We set the
normalisation of {\sc optxagnf} as equal to the magnification factor,
fix the soft X-ray excess to typical values of $kT_{e}=0.2$~keV and
an optical depth of $\tau=15$, fix the fraction of coronal emission to 
a typical value of $f_{pl}=0.3$, with spectral index $\Gamma=2.2$ as observed. 
We find that the broadband SED is well reproduced by a model with $\mu=6$, $M_{\rm BH}=1\times10^{10}$\msun, $L_{\rm bol}/L_{\rm Edd}=1$ and 
$r_{\rm corona}=25r_{\rm g}$.
This coronal radius is a typical value for AGN with an
accretion rate close to Eddington \citep{Jin2012a}. This intrinsic spectrum is 
shown by the red line marked 'Nominal' in 
Fig.\ref{fig:broadband}, whereas the magenta line in this figure 
includes the effect of absorption 
in our Galactic the cold absorber with
$N_{\rm H}=5\times 10^{22}$~cm$^{-2}$ introduced to explain the strong X-ray
continuum absorption. We only include dust reddening from 
our Galaxy because the very high column density in the cold absorber
would strongly suppress the optical flux down to much lower level than the observed
flux. This means the cold absorber is not dusty so it cannot be associated with the torus or
other material further out in the host galaxy. 

The magenta line also includes a torus template by \cite{Silva2004} to
reproduce the mid-IR data observed by {\it Spitzer}. We use this torus
template to estimate the power of dust emission, and its ratio to the
accretion power.  This ratio must be less than unity as the torus is
powered by reprocessing of the illuminating AGN flux. However, Table
\ref{tab:sedmodel} shows that the torus is more luminous than the
total accretion power in the Nominal SED model. This discrepancy is only
made worse if the magnification is different between the IR and the
nuclear region, as more extended IR will have smaller magnification.

We investigate the effect of changing the model parameters.  Firstly,
we investigate what happens with a larger black hole mass, as
reverberation mapping is clearly very difficult with the UV lines. A
higher black hole mass would give a lower temperature of the accretion
disc, but there is a limit to how low this can go while still fitting
the {\it INT} optical data. The model flux cannot be lower than this
data, but can be higher as there is substantial Lyman-alpha forest
absorption which additionally suppresses the optical/UV spectrum.
We find that this requirement means that there is an upper limit of the black
hole mass for zero spin of $M_{\rm BH}\sim2\times10^{10}$\msun, which is plotted
in green in Fig.~\ref{fig:broadband}, called 'high mass'. In this parameter set, 
we use an lower limit of
the Eddington ratio ($L_{\rm bol}/L_{\rm Edd}=0.5$) determined by the correlation
of photon index and Eddington ratio (see Sec.~\ref{sec:eddington}) since the lower
accretion rate decreases the disc temperature. The magnification factor is $\mu=4$,
and the corona radius is $r_{\rm corona}=40r_{\rm g}$, which is reasonable for a
half Eddington accretion, but now the mismatch with the IR emission is even worse
(see Table \ref{tab:sedmodel}). 

A possible solution to reproduce both the accretion flow emission and the torus reprocessing is 
to shift the peak of the quasar radiation to higher energies, into the unobservable far UV. 
The maximum disc temperature $T_{\rm max}$ follows a proportional relation
\begin{eqnarray}
T_{\rm max}\propto \dot{m}^{1/4}{M_{\rm BH}}^{-1/4}{r_{\rm ISCO}}^{-3/4}
\label{eq:Tmax}
\end{eqnarray}
so is higher for a larger Eddington ratio $\dot{m}$($\equiv L_{\rm bol}/L_{\rm Edd}$),
a smaller
black hole mass $M_{\rm BH}$ or a smaller disc inner radius $r_{\rm ISCO}$
(which depends on the black hole spin $a^*$). We have to simultaneously
reproduce the observed optical flux $F_{\rm opt}\propto\mu(M_{\rm BH}\dot{M}_{\rm acc})^{2/3}$ \citep{Davis2011}, with the additional relation from the bolometric luminosity $L_{\rm bol}\propto\dot{m}M_{\rm BH}\propto\eta\dot{M}_{\rm acc}$, where $\eta$ is the spin dependent efficiency of the energy conversion. This requires 
\begin{eqnarray}
\dot{m}\propto\eta \mu^{-3/2}{M_{\rm BH}}^{-2}. \label{eq:opt}
\end{eqnarray}

The solutions are limited by ranges of the parameters. The lower limit
of the black hole mass can be estimated by a comparison of the width
of the broad line with \pds.  \pds ($M_{\rm BH}=1\textrm{--}2\times
10^9$\msun) has an H$\beta$ line width of $3974\pm764$~km~s$^{-1}$
\citep{Torres1997}, a factor 2 smaller than an H$\alpha$ line width of
$7721$~km~s$^{-1}$ in \apm \citep{Oyabu2009}. The black hole mass
depends on the velocity and radius of the broad line region, with
$M_{\rm BH}\propto Rv^2$. Hence the black hole mass of \apm must be
larger than $4\times10^{9}$\msun even in the unlikely case that $R$ is
the same. While the upper limit of the magnification factor is
estimated by the reverberation mapping \citep{Saturni2016}, the lower
limit is not. Here, we simply set a lower limit to be $\mu=2$ since no
magnification $\mu=1$ seems to be unlikely for such a high luminosity
object. We consider two possible upper limits for the spin parameter
in this source.  We first assume that the spin is less than
$a^*\sim0.7$, as might perhaps be appropriate if this source is
radio-quiet, as high spin may always result in a powerful radio jet
\citep{Maraschi2012,Done2016}.

\begin{table}
\caption{Parameters for the solutions to explain the broadband SED}
\begin{center}
\begin{tabular}{lccccc}
\hline
\hline
& $M_{\rm BH}$ (\msun) & $a^*$ & $\dot{m}$ & $\mu$ & $\frac{L_{\rm dust}}{L_{\rm AGN}}$\\
\hline
Nominal (red) & $10^{10}$ & 0 & 1 & 6 & 1.35\\
High mass (green) & $2\times10^{10}$ & 0 & 0.5 & 4 & 2.03\\ 
Moderate spin (dark blue) & $10^{10}$ & 0.7 & 1 & 8 & 0.99\\
Maximum spin (blue) & $10^{10}$ & 0.998 & 1 & 15 & 0.48\\
Super-Eddington (orange) & $4\times10^{9}$ & 0 & 25 & 2 & 0.40\\
\hline
\end{tabular}
\end{center}
\label{tab:sedmodel}
\end{table}%

A solution with a moderate black hole spin of $a^*=0.7$ at Eddington 
is plotted in dark blue in Fig.~\ref{fig:broadband}. Higher spin gives higher efficiency so higher
luminosity for a given mass accretion rate through the outer disc. 
The magnification factor must be larger
to compensate the increase of $\eta$ following Eq.~\ref{eq:opt}. We find that settig the 
magnification factor at its upper limit ($\mu=8$) can reproduce the observed spectra.
In this case, the ratio of reprocessed IR to total 
accretion power is $\sim99\%$, which just barely avoids violating energy conservation. 

We also consider a maximally spinning black hole (spin parameter of
$a^*=0.998$).  This requires a large magnification factor $\mu=15$ for
$L_{\rm bol}/L_{\rm Edd}=1$ but the torus luminosity fraction is now a
more acceptable value of 48\%. While highly relativistic jets are not
well understood, it is clear that spin plays at least some role.  \apm
does have radio emission and its luminosity does put it into the FRII
category so it could indeed have a misaligned, highly powerful radio
jet, but some (perhaps all) of this radio emission is powered by the
strong star formation \citep{Riechers2009}.  The high resolution radio
images show no sign of a double radio structure \citep{Riechers2009},
but these could be suppressed by being on scales which are outside of
the lensing magnificantion and/or the source could be young, like the
GHz peaked AGN \citep{Bruni2015}.

The other possible solutions are higher accretion rate or/and lower
black hole mass.  A lower limit of black hole mass $M_{\rm
  BH}=4\times10^9$\msun corresponds to $\dot{m}=6.25$ by
Eq.~\ref{eq:opt} if the magnification factor and the spin are
unchanged. By using a lower limit of magnification factor $\mu=2$, the
Eddington ratio becomes as high as $\dot{m}=25$. This magnification
factor is slightly lower than that calculated from Eq.~\ref{eq:opt}
because the relation $F_{\rm opt}\propto\mu(M_{\rm BH}\dot{M}_{\rm
  acc})^{2/3}$ is only valid at energies below the disc peak.  The
model spectra with these parameters are plotted in orange in
Fig.~\ref{fig:broadband}, again giving a more acceptable luminosity
fraction of the torus compared to the accretion power of 40\%.

Exactly where the SED peaks is also important for the launch mechanism
of the wind.  According to the discussion in \cite{Laor2014}, the
UV-line driven disc wind is efficiently accelerated by radiation with
a effective temperature of $\sim30000\textrm{--} 50000$~K. This
corresponds to a peak in $\nu f_\nu$ at $\sim0.01$~keV. Both nominal
and high mass models peak at energies which are somewhat below this,
so may not have enough UV for efficient line driving. However, these
are the two models which were ruled out by energy conservation as they
cannot power the observed IR radiation. Instead, the 
higher spin and super-Eddington models all have 
copious UV photons at $\sim0.01$~keV, yet are also X-ray weak. 
Thus, the 
broad band SED of \apm is like that of \pds, in being UV bright and X-ray weak,
as required for efficient UV line driving. 

\subsection{Relation with the broad UV absorption}
\begin{figure}
\begin{center}
\includegraphics[width=\hsize]{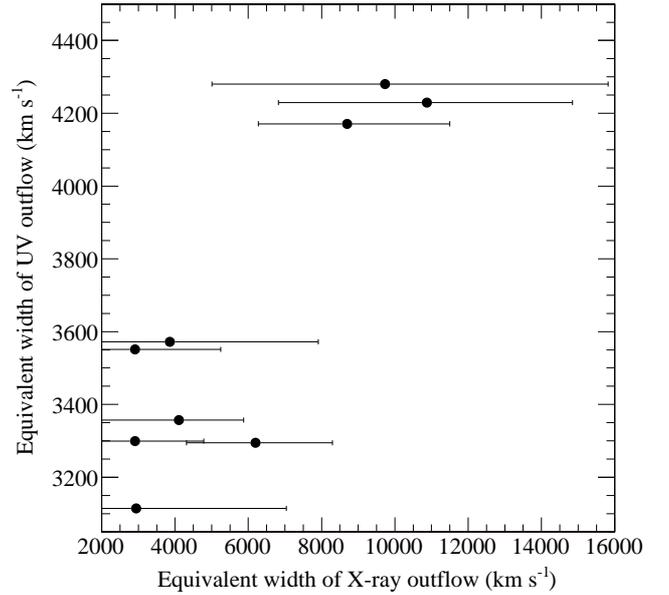}
\caption{
The equivalent width of the UV outflows (co-added BAL and blue and red NALs) versus that of X-ray outflow.
The decrease in X-ray equivalent width with time seen from our data (modelled as a change in viewing angle: 
lower panel of Fig. 5) is correlated with the 
decrease in the UV absorption seen by \citet{Trevese2013}. 
}
\label{fig:EWcorrelation}
\end{center}
\end{figure}

Spectral fitting with our hot wind with cool clumps model revealed that the outflow
velocity decreases from $\sim0.2c$ to $\sim0.1c$ while the viewing angle (which traces the
column density) of the wind also decreases from $\sim48^\circ$ to $\sim46^\circ$
during six years between 2001-10-30 and 2008-01-14. This was not discussed in
previous work because their analysis concentrated on the putative faster components
of the wind. 

Similar trends are observed in the outflows seen in UV band. \cite{Trevese2013}
investigated the long term variability of the absorption line profiles from 1998 to 2012,
which spans the time of the X-ray observations.
They separate the \civ absorption systems into 4 components, two which are blends of
narrower lines (blue NAL and red NAL) and two which are intrinsically broad
(BAL1 and BAL2). They found that the equivalent
widths of all these UV features decrease together, so we coadd their equivalent widths
to measure the total UV absorption.
We estimate this at the time of the X-ray observations by a linear interpolation.

The X-ray equivalent width was evaluated from the viewing angle of our
wind model (see Fig.~\ref{fig:angle}), and converted into units of
velocity km~s$^{-1}$ for consistency with the values of the UV
outflows.  Fig.~\ref{fig:EWcorrelation} shows that the X-ray and UV
absorption equivalent widths decrease together. This would be easy to
explain if the SED has changed as then both winds would be responding
to the same change in ionising flux. However, \cite{Trevese2013} shows
the R-band magnitude is stable within less than $\sim10$\%, and the
X-ray intrinsic luminosity is also stable to within $\sim50\%$ as
shown in Sec.~\ref{sec:eddington}.

Instead, we could explain the correlated change if both absorbers are
a part of the same structure. However, this seems most unlikely as
they have very different velocities, 0.1--0.2$c$ for the X-ray
absorber and $0.04c$ for the UV, and very different velocity behaviour
as a function of time. The X-ray absorber slows (see Table 4) but UV absorption
systems remain at a fixed velocity.

It is possible that the X-ray wind acts as a shield for the UV wind
\citep{Murray1998}. A decreasing column density of the X-ray wind would then
increase irradiation of the UV wind, decreasing the \civ ion
fraction. We estimate the time scale on which the column density can change
considering the large black hole mass and the high redshift of this
source.  The wind in \apm changes over 6 years, which corresponds to 1
year at the quasar's rest frame. This corresponds to a size scale of
$\sim100r_{\rm g}$ assuming a mass of $10^{10}$\msun using a velocity
of $\sim0.1\textrm{--}0.2c$. This scale is very similar to the
inhomogeneity of the UV-line driven disc wind seen in the hydrodynamic
simulations \citep[e.g.,][]{Proga2000}.  In the archetypal wind source
\pds, the time scale for this kind of change is much smaller. Since
the black hole mass is $\sim10$ times larger, and time dilation by the
redshift is $\sim5$ times larger than \pds, the typical time scale in
\pds is $\sim50$ times smaller than \apm. It means that the six-year
decrease of the mass outflow rate in this source corresponds to a
variability in $\sim1.5$ months in \pds.

\subsection{The other gravitationally lensed quasars}

A few gravitationally lensed quasars at high redshift have features around iron which have
been  modelled by extremely
smeared reflection \citep{Reis2014,Reynolds2014,Walton2015,Lanzuisi2016}. These fits 
result in a high black hole spin of $a^{*}\sim0.7\textrm{--}0.9$, which has important implications for
black hole evolution across cosmic time. 
However, the iron-K spectral feature in these sources might be interpreted as being instead from 
a wind, similar to that seen in \apm. \cite{Lanzuisi2016} show that partially ionised, partial covering
matches the spectral features in the lensed redshift $\sim2$ quasar PG~1247+267
as well (or even better than) relativistic reflection models.
As shown in \cite{Hagino2016}, the wind absorption with
a larger viewing angle produces a similar spectral feature to the extremely smeared
relativistic reflection. As an example, the iron-K feature in the most extreme relativistic reflection
source \hh is successfully explained by the wind model.

\section{Conclusions}
We have successfully explained the observed X-ray spectra of \apm by a
hot disc wind and a cool partially covering absorber with a
non-extreme wind velocity $\sim0.1\textrm{--}0.2c$. This model does not 
require the extremely fast disc wind of  previous work
because the higher energy absorption feature is matched by the
additional curvature from the edge structure in the cooler material
rather than by an extremely broadened absorption line. This means that
the wind in our interpretation can be powered by radiation driving
especially as the source is UV bright, X-ray weak and close to (or
exceeding) the Eddington limit. However, to unambiguously distinguish
our model and the extreme wind model requires much better data than
currently available.

We show that the X-ray absorber has decreased in both velocity and
equivalent width over a timespan of 6 years, correlated with a
decrease in the equivalent width (but not velocity) of the $\sim0.04c$
\civ UV Broad Absorption Line. This correlation is not driven by the
observed illuminating flux as this remains constant. Instead, it may
indicate that the X-ray wind acts as a shield for the BAL.

\section*{ACKNOWLEDGMENTS}
The authors are grateful to Martin~Hardcastle for discussions on the radio emission
from \apm.
K.H. was supported by the Japan Society for the
Promotion of Science (JSPS) Research Fellowship for Young
Scientists.
This work was supported by JSPS KAKENHI grant numbers 15H06897 and 24105007.
CD acknowledges STFC funding under grant ST/L00075X/1 and a JSPS long term fellowship.

\bibliographystyle{mnras}
\bibliography{ref}
 
\bsp    
\label{lastpage}
\end{document}